\documentclass[twocolumn]{aastex63}

\usepackage{subfigure}

\newcommand{\ktwo}{K2}

\newcommand{\rearth}{R$_\oplus$}
\newcommand{\mearth}{M$_\oplus$}

\newcommand{\hd}{HD 106315}
\newcommand{\gj}{GJ 9827}
\newcommand{\hdb}{HD 106315 b}
\newcommand{\hdc}{HD 106315 c}
\newcommand{\gjb}{GJ 9827 b}
\newcommand{\gjc}{GJ 9827 c}
\newcommand{\gjd}{GJ 9827 d}

\shorttitle{HD 106315 \& GJ 9827}
\shortauthors{Kosiarek et al. 2020}
\graphicspath{{./}{figures/}}

\begin{document}

\title{Physical Parameters of the Multi-Planet Systems HD 106315 and GJ 9827\footnote{Based on observations obtained at the W. M. Keck Observatory, which is operated jointly by the University of California and the
California Institute of Technology.}\footnote{This paper includes data gathered with the 6.5 meter Magellan Telescopes located at Las Campanas Observatory, Chile.}}

\correspondingauthor{Molly R.\ Kosiarek}
\email{molly.kosiarek@gmail.com}

\author[0000-0002-6115-4359]{Molly R.\ Kosiarek}
\altaffiliation{NSF Graduate Research Fellow}
\affiliation{Department of Astronomy and Astrophysics, University of California, Santa Cruz, CA 95064, USA}

\author[0000-0001-6298-412X]{David A. Berardo}
\affiliation{Department of Physics, and Kavli Institute for Astrophysics and Space Research, Massachusetts Institute of Technology, Cambridge, MA 02139, USA}

\author{Ian J. M. Crossfield}
\affiliation{Department of Physics \& Astronomy, University of Kansas, 1082 Malott,1251 Wescoe Hall Dr., Lawrence, KS 66045, USA}

\author[0000-0003-1314-7514]{Cesar Laguna}
\affiliation{Department of Physics, University of California, Santa Cruz, CA 95064, USA}

\author[0000-0002-2875-917X]{Caroline Piaulet}
\affil{Department of Physics, and Institute for Research on Exoplanets, Universit\'{e} de Montr\'{e}al, Montreal, H3T 1J4, Canada}

\author[0000-0001-8898-8284]{Joseph M. Akana Murphy}
\altaffiliation{NSF Graduate Research Fellow}
\affiliation{Department of Astronomy and Astrophysics, University of California, Santa Cruz, CA 95064, USA}

\author[0000-0002-2532-2853]{Steve~B.~Howell}
\affil{NASA Ames Research Center, Moffett Field, CA 94035, USA}

\author{Gregory W. Henry}
\affil{Center of Excellence in Information Systems, Tennessee State University, Nashville, TN  37209, USA}

\author[0000-0002-0531-1073]{Howard Isaacson}
\affiliation{501 Campbell Hall, University of California at Berkeley, Berkeley, CA 94720, USA}
\affiliation{Centre for Astrophysics, University of Southern Queensland, Toowoomba, QLD, Australia}

\author[0000-0003-3504-5316]{Benjamin Fulton}
\affiliation{NASA Exoplanet Science Institute/Caltech-IPAC, MC 314-6, 1200 E California Blvd, Pasadena, CA 91125, USA}

\author[0000-0002-3725-3058]{Lauren M. Weiss}
\affiliation{Institute for Astronomy, University of Hawai`i, 2680 Woodlawn Drive, Honolulu, HI 96822, USA}

\author[0000-0003-0967-2893]{Erik A. Petigura}
\affiliation{Department of Physics \& Astronomy, University of California Los Angeles, Los Angeles, CA 90095, USA}

\author[0000-0003-0012-9093]{Aida Behmard}
\affiliation{Division of Geological and Planetary Sciences, California Institute of Technology, Pasadena, CA 91125, USA}

\author[0000-0001-8058-7443]{Lea A. Hirsch}
\affiliation{Kavli Institute for Particle Astrophysics and Cosmology, Stanford University, Stanford, CA, USA}

\author{Johanna Teske}
\affil{Earth \& Planets Laboratory, Carnegie Institution of Washington 5241 Broad Branch Road, N.W., Washington, DC 20015, USA}

\author[0000-0002-0040-6815]{Jennifer A. Burt} 
\affiliation{Jet Propulsion Laboratory, California Institute of Technology, 4800 Oak Grove Drive, Pasadena, CA 91109, USA}

\author[0000-0002-4535-6241]{Sean M. Mills} 
\affiliation{Department of Astronomy, California Institute of Technology, Pasadena, CA 91125, USA}

\author[0000-0003-1125-2564]{Ashley Chontos}
\altaffiliation{NSF Graduate Research Fellow} 
\affiliation{Institute for Astronomy, University of Hawai‘i, Honolulu, HI 96822, USA}


v\author[0000-0003-4603-556X]{Teo Mo\v{c}nik}
\affiliation{Gemini Observatory/NSF's NOIRLab, 670 N. A'ohoku Place, Hilo, HI 96720, USA}

\author[0000-0001-8638-0320]{Andrew W.\ Howard}
\affiliation{Department of Astronomy, California Institute of Technology, Pasadena, CA 91125, USA}

\author[0000-0003-4990-189X]{Michael Werner}
\affiliation{Jet Propulsion Laboratory, California Institute of Technology, 4800 Oak Grove Drive, Pasadena, CA 91109, USA}

\author[0000-0002-4881-3620]{John~H.~Livingston}
\affiliation{Department of Astronomy, University of Tokyo, 7-3-1 Hongo, Bunkyo-ku, Tokyo 113-0033, Japan}

\author[0000-0002-2413-5976]{Jessica Krick}
\affiliation{Caltech/IPAC, 1200 E. California Blvd. Pasadena, CA 91125}

\author[0000-0002-5627-5471]{Charles Beichman}
\affiliation{IPAC/NASA Exoplanet Science Institute, Caltech, Jet Propulsion Laboratory}

\author[0000-0002-8990-2101]{Varoujan Gorjian}
\affiliation{Jet Propulsion Laboratory, California Institute of Technology, 4800 Oak Grove Drive, Pasadena, CA 91109, USA}

\author[0000-0003-0514-1147]{Laura Kreidberg}
\affiliation{Max Planck Institute for Astronomy, K{\"o}nigstuhl 17, 69117 Heidelberg, Germany}
\affiliation{Center for Astrophysics $|$ Harvard \& Smithsonian, 60 Garden Street, Cambridge, MA, 02138, USA}

\author[0000-0002-4404-0456]{Caroline Morley}
\affiliation{University of Texas at Austin, Austin, TX, 78712, USA}

\author[0000-0001-9414-3851]{Jessie L. Christiansen}
\affiliation{Caltech/IPAC, 1200 E. California Blvd. Pasadena, CA 91125}

\author[0000-0001-9414-3851]{Farisa Y. Morales}
\affiliation{Jet Propulsion Laboratory, California Institute of Technology, 4800 Oak Grove Drive, Pasadena, CA 91109, USA}

\author[0000-0003-1038-9702]{Nicholas J. Scott}
\affil{NASA Ames Research Center, Moffett Field, CA 94035, USA}

\author[0000-0002-5226-787X]{Jeffrey D. Crane}
\affiliation{The Observatories of the Carnegie Institution for Science 813 Santa Barbara Street, Pasadena, CA, USA 91107}

\author[0000-0002-6937-9034]{Sharon Xuesong Wang}
\affiliation{Observatories of the Carnegie Institution for Science, 813 Santa Barbara St., Pasadena, CA 91101}
\affiliation{Department of Astronomy, Tsinghua University, Beijing 100084, People's Republic of China}

\author{Stephen A. Shectman}
\affiliation{The Observatories of the Carnegie Institution for Science 813 Santa Barbara Street, Pasadena, CA, USA 91107}

\author{Lee J. Rosenthal}
\affiliation{Department of Astronomy, California Institute of Technology, Pasadena, CA 91125, USA}

\author[0000-0003-4976-9980]{Samuel K. Grunblatt}
\affiliation{American Museum of Natural History, 200 Central Park West, Manhattan, NY 10024, USA}
\affiliation{Center for Computational Astrophysics, Flatiron Institute, 162 5$^{\rm{th}}$ Avenue, Manhattan, NY 10010, USA}

\author[0000-0003-3856-3143]{Ryan A. Rubenzahl}
\altaffiliation{NSF Graduate Research Fellow}
\affiliation{Department of Astronomy, California Institute of Technology, Pasadena, CA 91125, USA}

\author[0000-0002-4297-5506]{Paul A. Dalba}
\altaffiliation{NSF Astronomy and Astrophysics Postdoctoral Fellow}
\affiliation{Department of Earth and Planetary Sciences, University of California, Riverside, CA 92521, USA}

\author[0000-0002-8965-3969]{Steven Giacalone}
\affil{Department of Astronomy, University of California Berkeley, Berkeley, CA 94720-3411, USA}

\author{Chiara Dane Villanueva}
\affiliation{Department of Physics, University of California, Santa Cruz, CA 95064, USA}

\author{Qingtian Liu}
\affiliation{Department of Physics, University of California, Santa Cruz, CA 95064, USA}

\author[0000-0002-8958-0683]{Fei Dai}
\affiliation{Division of Geological and Planetary Sciences, California Institute of Technology, Pasadena, CA 91125, USA} 

\author[0000-0002-0139-4756]{Michelle L. Hill}
\affiliation{Department of Earth and Planetary Sciences, University of California, Riverside, CA 92521, USA}

\author[0000-0002-7670-670X]{Malena Rice}
\affiliation{Department of Astronomy, Yale University, New Haven, CT 06511, USA}

\author[0000-0002-7084-0529]{Stephen R. Kane}
\affiliation{Department of Earth and Planetary Sciences, University of California, Riverside, CA 92521, USA}

\author[0000-0002-7216-2135]{Andrew W. Mayo}
\affiliation{Department of Astronomy, University of California Berkeley, Berkeley, CA 94720-3411, USA}
\affiliation{Centre for Star and Planet Formation, Natural History Museum of Denmark \& Niels Bohr Institute, University of Copenhagen, Øster Voldgade 5-7, DK-1350 Copenhagen K., Denmark}

\nocollaboration{50}



\begin{abstract}

\hd\ and \gj\ are two bright, nearby stars that host multiple super-Earths and sub-Neptunes discovered by K2 that are well suited for atmospheric characterization.
We refined the planets' ephemerides through Spitzer transits,
enabling accurate transit prediction required for future atmospheric characterization through transmission spectroscopy. 
Through a multi-year high-cadence observing campaign with Keck/HIRES and Magellan/PFS, we improved the planets' mass measurements in anticipation of Hubble Space Telescope transmission spectroscopy. 
For \gj, we modeled activity-induced radial velocity signals with a Gaussian process informed by the Calcium II H\&K lines in order to more accurately model the effect of stellar noise on our data. 
We measured planet masses of M$_b$=$4.87\pm 0.37$ \mearth, M$_c$=$1.92\pm 0.49$ \mearth, and M$_d$=$3.42\pm 0.62$ \mearth. For \hd,
we found that such activity-radial velocity decorrelation was not effective due to the reduced presence of spots and speculate that this may extend to other hot stars as well (T$_{\rm {eff}}>6200$ K).   
We measured planet masses of M$_b$=$10.5\pm 3.1$ \mearth\ and M$_c$=$12.0\pm 3.8$ \mearth.
We investigated all of the planets' compositions through comparing their masses and radii to a range of interior models. \gjb\ and \gjc\ are both consistent with a 50/50 rock-iron composition, \gjd\ and \hdb\ both require additional volatiles and are consistent with moderate amounts of water or hydrogen/helium, and \hdc\ is consistent with a $\sim$10\% hydrogen/helium envelope surrounding an Earth-like rock \& iron core.
\end{abstract}


\section{Introduction} \label{sec:intro}
Small planets cover a wide variety of compositions ranging from dense, iron-rich planets to low density planets with large hydrogen/helium envelopes. 
Mass and radius are degenerate with many potential compositions; measurements of atmospheric compositions can help break this degeneracy \citep{Figueira2009,Rogers2010}. 

In this paper, we characterize two planetary systems, HD 106315 and GJ 9827. These systems both consist of multiple planets transiting bright, nearby host stars.
Both systems contain promising targets for atmospheric composition studies through transmission spectroscopy. Three of the planets are being observed by the Hubble Space Telescope (HST) to study their atmospheres in GO-15333 (\citealp{Kreidberg2020}, Benneke et al. in prep) and GO-15428 (Hedges et al. in prep). 
These three planets are additionally compelling targets for the James Webb Space Telescope (JWST) as determined by their transmission spectroscopy metric values \citep[TSM,][\hdc: 91, \gjb: 95, \gjd: 144]{Kempton2018}. Precise mass measurements (20\% precision) are needed to support the ongoing HST analyses and potential JWST observations as mass directly affects the observability of features and inferred properties from spectra \citep{Batalha2019}.

We measure the planet radii and update their ephemerides with Spitzer transit observations in Section~\ref{sec:spitzer}. We describe our spectroscopy, imaging data, and update stellar parameters in Section~\ref{sec:stellarchar}. 
We investigate stellar activity in our radial velocity observations, K2 photometry, and ground-based photometry in Section~\ref{sec:stellaractivity}. We refine the planet masses through radial velocity analyses and explore the stability of including non-zero eccentricities with N-body simulations in Section~\ref{sec:rv}. Finally, we examine potential interior compositions in Section~\ref{sec:interior} by comparing the masses and radii with composition models, before concluding in Section~\ref{sec:conclusion}.

\subsection{GJ 9827}

GJ 9827 (K2-135) is a bright (V=10.3 mag, K=7.2 mag), nearby (distance=30 pc) K6 dwarf star hosting three planets discovered in K2 Campaign 12 \citep{Niraula2017,Rodriguez2018}. Planets b and c orbit near a 3:1 resonance at 1.2 days and 3.6 days, with planet d at 6.2 days. These three planets span the gap seen in the radius distribution of small planets \citep{Fulton2017} sized at 1.529$\pm$0.058 \rearth, 1.201$\pm$0.046 \rearth, and 1.955$\pm$0.075 \rearth\ respectively. \citet{Niraula2017} additionally collected 7 radial velocity observations with the FIbrefed Echelle Spectrograph \citep[FIES;][]{Frandsen1999,Telting2014} to vet the system and to derive stellar parameters. 

The mass of planet b was first determined with radial velocity observations from the Carnegie Planet Finder Spectrograph \citep[PFS,][]{Crane2006,Crane2008,Crane2010} on Magellan II by \citet{Teske2018} (M$_b\sim$8 \mearth), who placed upper limits on planets c and d (M$_c<$2.5 \mearth, M$_d<$5.6 \mearth). Through additional measurements with the High Accuracy Radial velocity Planet Searcher (HARPS, \citealp{Mayor2003}) 
and the High Accuracy Radial velocity Planet Searcher for the Northern hemisphere (HARPS-N), 
\citet{PrietoArranz2018} determined the masses of all three planets (M$_b$=3.74$\pm$0.50 \mearth, M$_c$=1.47$\pm$0.59 \mearth, and M$_d$=2.38$\pm$0.71 \mearth). 
The masses of planets b and d were further refined by \citet{Rice2019} with new HARPS-N radial velocity measurements and a Gaussian process informed by the K2 light curve  (M$_b$=4.91$\pm$0.49 \mearth\ and M$_d$=4.04$\pm$0.84 \mearth). Both \citet{PrietoArranz2018} and \citet{Rice2019} discuss how the inner planets have a high density and the outer planet has a lower density, suggesting that photoevaporation or migration could have played a role in the evolution of this system; we discuss this possibility further in Section~\ref{sec:interior}. 



\subsection{HD 106315}

HD 106315 (K2-109) is a bright (V=8.97 mag, K=7.85 mag) F5 dwarf star hosting two planets discovered in K2 Campaign 10 \citep{Crossfield2017, Rodriguez2017}.
Planet b is a small (R$_b$=${2.40\pm0.20}$ \rearth) planet with an orbital period of 9.55 days; planet c is a warm Neptune-sized (R$_c$=${4.379\pm0.086}$ \rearth) planet with an orbital period of 21.06 days.

This system was further characterized with HARPS radial velocity observations by \citet{Barros2017} to determine the planets' masses ($M_b$=$12.6 \pm3.2$ \mearth\ and $M_c$=$15.2\pm3.7$ \mearth). They concluded that \hdb\ likely has a rocky core and decent water mass fraction whereas \hdc\ has a substantial hydrogen-helium envelope. 

Additional transits of \hdc\ were observed with two ground based facilities: the Euler telescope \citep{Lendl2017} and the Cerro Tololo Inter-American Observatory \citep[CTIO,][]{Barros2017}. These measurements improved the precision on both the orbital period and the time of transit.

Later \citet{Zhou2018} investigated the system architecture through measuring the obliquity of \hdc\ using Doppler tomography and constraining the mutual inclination of \hdb\ through dynamical arguments. They found that these two planets both have low obliquities,
consistent with the few other warm Neptunes with measured obliquities \citep[eg.][]{Albrecht2013}.

\section{Spitzer Transits}
\label{sec:spitzer}

Predicting precise future transit times becomes harder as more time elapses from previous transit observations and the uncertainty from the orbital period compounds. These systems contain promising targets for future atmospheric follow-up which require small uncertainties on the predicted transit time. Therefore, we collected additional transit observations on the Spitzer Space Telescope to refine the ephemerides for each planet as well as to provide a depth measurement at 4.5$\mu m$. 
These observations were taken as part of the K2 follow-up program 13052 (PI: Werner), using the 4.5$\mu$m channel of IRAC \citep{Fazio2004}. A single transit of each planet was observed, except for \hdb\ which was observed twice. All of the observations were collected with 0.4 second exposures and the target placed on the `sweet spot' of the detector. 

\begin{figure}[tbp]
\centering
\includegraphics[trim={0 8cm 0 8cm},clip, width=0.5\textwidth]{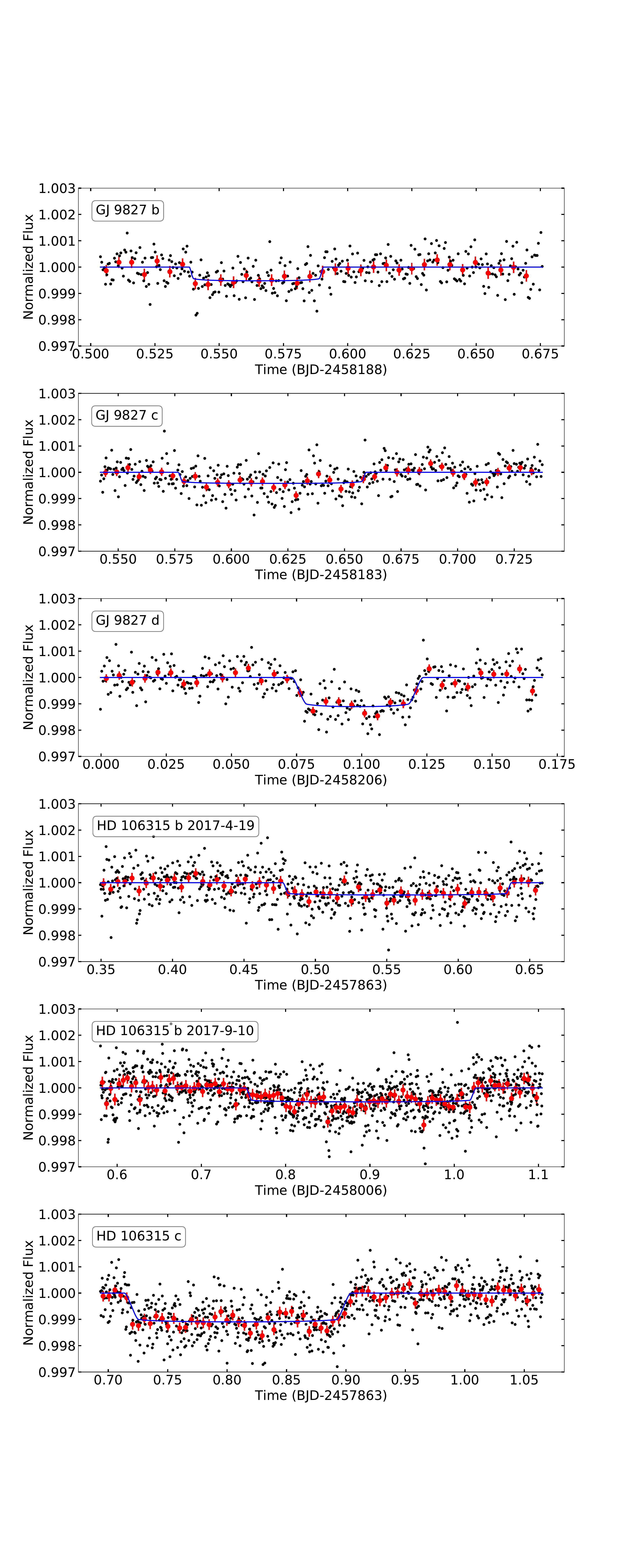} 
\caption{Spitzer transits for GJ 9827 b, c, d and HD 106315 b, c. Data (black points), binned data (red circles), and model fit (blue line) are shown.
\label{fig:spitzer}}

\end{figure}

\begin{deluxetable*}{rcccccc}[tbp]
\tablecaption{Spitzer Transit Results\label{tab:sp}}
\tablehead{\colhead{Planet} & \colhead{Date (UT)} & \colhead{Time of Conjunction (BJD)} & \colhead{Rp/R$_*$ (4.5$\micron$)} & \colhead{Semimajor Axis (R$_*$)} & \colhead{Inclination ($^{\circ}$)} & \colhead{Uncertainty (dex)}} 
\startdata
GJ 9827 b & 2018-03-10 & 2457738.82384$^{+0.00081}_{-0.00080}$ & 0.0225$^{+0.0018}_{-0.0017}$ & 7.19$^{+0.56}_{-0.40}$ & 87.7$^{+1.8}_{-1.6}$ & -3.152$^{+0.012}_{-0.012}$ \\
GJ 9827 c & 2018-03-06 & 2457742.1993$^{+0.0025}_{-0.0028}$ & 0.0201$^{+0.0023}_{-0.0020}$ & 13.0$^{+1.7}_{-1.3}$ & 88.5$^{+1.4}_{-1.1}$ & -3.307$^{+0.015}_{-0.017}$  \\
GJ 9827 d & 2018-03-28 & 2457740.98800$^{+0.00064}_{-0.00055}$ & 0.0348$^{+0.0014}_{-0.0013}$ & 21.8$^{+2.5}_{-1.6}$ & 87.72$^{+0.37}_{-0.21}$ & -3.295$^{+0.017}_{-0.018}$  \\
HD 106315 b & 2017-4-19  & 2457586.5394$^{+0.0056}_{-0.0109}$ & 0.0201$^{+0.0026}_{-0.0024}$ & 16.4$^{+5.1}_{-3.1}$ & 88.4$^{+2.3}_{-1.1}$ & -3.197$^{+0.013}_{-0.013}$  \\
HD 106315 b & 2017-9-10 & 2457586.5826$^{+0.0121}_{-0.0043}$ & 0.0219$^{+0.0034}_{-0.0026}$ & 10.4$^{+2.2}_{-1.3}$ & 87.6$^{+3.0}_{-1.7}$ & -3.155$^{+0.010}_{-0.010}$  \\
HD 106315 c & 2017-4-20 & 2457569.0103$^{+0.0012}_{-0.0012}$ & 0.0329$^{+0.0013}_{-0.0012}$ & 29.5$^{+5.7}_{-4.2}$ & 88.89$^{+0.69}_{-0.51}$ & -3.189$^{+0.012}_{-0.012}$  
\enddata
\end{deluxetable*}

We follow a similar analysis approach to that described in \cite{Berardo2019}, which detrends the data using the Pixel Level Decorrelation method outlined in  \cite{Deming2015}. In brief, we first applied a median filter to each pixel in the image and calculated a background level for each frame by taking the median of the flux in an annulus centered on the point spread function. We estimated the centroid of each frame by fitting a two dimensional Gaussian to the image, and obtained a light curve using a fixed radius aperture. We varied the aperture size and performed a linear regression 
to determine the optimal radius; we found 2.4 pixels minimized the root mean square (RMS) of the residuals for all observations.

We modeled systematics in the light curve by weighting the nine brightest pixels individually as well as fitting for a quadratic time ramp. We then chose the combination of pixel coefficients, aperture size, and time-series binning that resulted in the smallest RMS deviation. We ran a Markov-Chain Monte Carlo (MCMC) analysis to estimate parameter uncertainties, using the systematic model in addition to a transit signal which we modeled using \texttt{batman} \citep{Kreidberg2015}. We fixed the period of each planet to the most recent measurements \citep{Barros2017,Rice2019}
and allowed the transit depth, center, orbital inclination, and semi major axis to vary. We also left the uncertainty of the data points as a free parameter, which we found converged to the RMS scatter of the raw light curve. We held fixed the quadratic limb darkening parameters, which were determined using the tables of \citet{Claret2011}. 
The fit results are shown in Table~\ref{tab:sp} and Figure~\ref{fig:spitzer}.

We calculated updated ephemerides (Table~\ref{tab:eph}) to further refine the time of conjunction and orbital period for future atmospheric follow-up and to better constrain these values in our radial velocity fits (Section~\ref{sec:rv}). We fit a straight line to the transit centers obtained from each individual observation, incorporating all ground-based published transits thus far \citep{Lendl2017,Barros2017}. 
These planets will be accessible for future transmission spectroscopy observations throughout the JWST era. As an example, the transit time uncertainty in 2025 is under two hours for all five planets (\gjb: 0.1hr, \gjc: 0.5hr, \gjd: 0.1hr, \hdb: 1.7hr, \hdc: 0.4hr).

\begin{deluxetable}{rcc}[btp]
\tablecaption{Ephemerides Update\label{tab:eph}}
\tablehead{\colhead{Planet} & \colhead{Time of Conjunction (BJD)} & \colhead{Period (days)}}
\startdata
GJ 9827 b & 2457738.82586$\pm$0.00026 & 1.2089765$\pm$2.3e-06 \\
GJ 9827 c & 2457742.19931$\pm$0.00071 & 3.648096$\pm$2.4e-05  \\
GJ 9827 d & 2457740.96114$\pm$0.00044 & 6.20183$\pm$1.0e-05  \\
HD 106315 b &  2457586.5476$\pm$0.0025 & 9.55287$\pm$0.00021 \\
HD 106315 c &   2457569.01767$\pm$0.00097 & 21.05652$\pm$0.00012
\enddata
\end{deluxetable}

\section{Stellar Parameters and Companion Refinement} 
\label{sec:stellarchar}

\subsection{Spectroscopy}
\label{sec:spectra}

We collected radial velocity measurements of \gj\ and \hd\ with the High Resolution Echelle Spectrometer (HIRES, \citealp{Vogt1994}) on the Keck I Telescope on Maunakea. These exposures were taken through an iodine cell for wavelength calibration \citep{Butler1996}. The HIRES data collection, reduction, and analysis followed the California Planet Search method described in \citet{Howard2010}. 

We obtained 92 measurements of \gj\ with HIRES between 2017 September 22 and 2020 January 8 (Table~\ref{tab:gjrvs}). These data were collected with the C2 decker (14\arcsec x0.861\arcsec, resolution=50k) with a typical signal-to-noise radio (SNR) of 200/pixel (250k on the exposure meter, median exposure time of 18.5 minutes). We also collected a higher resolution template observation with the B3 decker (14\arcsec x0.574\arcsec, resolution=67k) on 2017 December 30 with a SNR of 200/pixel without the iodine cell. Both the C2 and B3 decker allow for sky subtraction which is important for the quality of the radial velocities for a V=10 mag star. We included an additional 142 measurements in our \gj\ analysis, for a total of 234 measurements: 7 from FIES \citep{Niraula2017}, 36 from PFS \citep{Teske2018}, 35 from HARPS \citep{PrietoArranz2018}, and 64 from HARPS-N \citep{PrietoArranz2018,Rice2019}.

We obtained 352 measurements of \hd\ with HIRES between 2016 December 23 and 2020 Febuary 1 (Table~\ref{tab:hdrvs}); 53 of these observations were previously published in \citet{Crossfield2017}. These data were collected with the B5 decker (3.5\arcsec x0.861\arcsec, resolution=50k) with a typical SNR of 200/pixel (250k on the exposure meter, median exposure time of 4.8 minutes). Data were typically taken in groups of three consecutive observations to mitigate p-mode oscillations; \cite{Barros2017} estimated p-mode periods of $\sim$20 minutes whereas \citet{Chaplin2019} estimates timescales to be $\sim$30 minutes.
When possible, multiple visits separated by an hour were taken to improve precision due to the high $v \sin i$; these data were then binned in nightly bins to average over short-timescale activity.
We also collected a higher resolution template observation with the B3 decker on 2016 December 24. The template was a triple exposure with a total SNR of 346/pixel (250k each on the exposure meter) without the iodine cell.

We obtained 25 measurements of \hd\ with PFS between 2017 January 6 and 2018 June 30 (Table~\ref{tab:hdrvs}). Data taken prior to 2018 February were taken with the 0.5\arcsec slit (resolution$\sim$80k); a single observation with an exposure time of 10 to 25 minutes was taken per night. After a PFS upgrade in 2018 February, multiple exposures were taken with the 0.3\arcsec slit (resolution$\sim$130k). As with the HIRES data, we binned these consecutive observations for our analysis. An iodine-free template, consisting of three 1000s exposures, was taken with the 0.3\arcsec slit on 2018 June 27. The PFS data were reduced using a custom IDL pipeline and velocities extracted based on the methodology described in \citet{Butler1996}.

We additionally include 84 measurements from HARPS 
\citep{Barros2017}, for a total of 461 measurements (160 binned points) in our \hd\ analysis.
We collected 125 measurements on the Automated Planet Finder \citep[APF,][]{Radovan2014,Vogt2014} but do not include them in the analysis due to the high scatter (30 m/s nightly RMS, 7.3 m/s RV uncertainty), listed in Table~\ref{tab:hdrvs}.

We updated the stellar parameters for \gj\ and \hd\ to incorporate the latest measurements, especially the Gaia DR2 parallaxes \citep{Gaia2016,Gaia2018,Luri2018}. We used multiband stellar photometry (Gaia G and 2MASS JHK), the Gaia parallax, and a stellar effective temperature and metallicity derived from Keck/HIRES spectra via the SpecMatch-Emp tool \citep{Yee2017}. The SpecMatch-Emp values are T$_\mathrm{eff}=6318 \pm 110$~K and $4195\pm70$~K, and [Fe/H]$=  -0.21 \pm 0.09$ and $-0.29 \pm 0.09$ for HD 106315 and GJ 9827, respectively. We input the above values into the isoclassify tool using the grid-mode option \citep{Huber2017} to derive the stellar parameters listed in Table~\ref{tab:stparams}.

\begin{deluxetable}{rccc}[h]
\tablecaption{Stellar Parameters \label{tab:stparams}}
\tablehead{\colhead{Parameter} & \colhead{units} & \colhead{\gj} & \colhead{\hd}}
\startdata
[Fe/H] & dex & -0.26$\pm$0.08 & -0.22$\pm$0.09 \\
M$_{*}$ & M$_{\rm Sun}$ & 0.593$\pm$0.018 & 1.154$\pm$0.042 \\
R$_{*}$ & R$_{\rm Sun}$ & 0.579$\pm$0.018 & 1.269$\pm$0.024 \\
log g & dex & 4.682$\pm$0.021 & 4.291$\pm$0.025 \\
T$_{\rm eff}$ & K & 4294$\pm$52 & 6364$\pm$87 
\enddata
\end{deluxetable}

\subsection{HD 106315 Imaging}

The discovery papers for \hd\ included seeing limited imaging data and K-band Keck/NIRC2 infrared adaptive optics imaging to rule out nearby stellar companions \citep{Rodriguez2017,Crossfield2017}. We include here additional high contrast imaging data to improve the magnitude contrast constraints on nearby companions.

\begin{figure}[htbp]
\begin{center}
\includegraphics[width=0.48\textwidth]{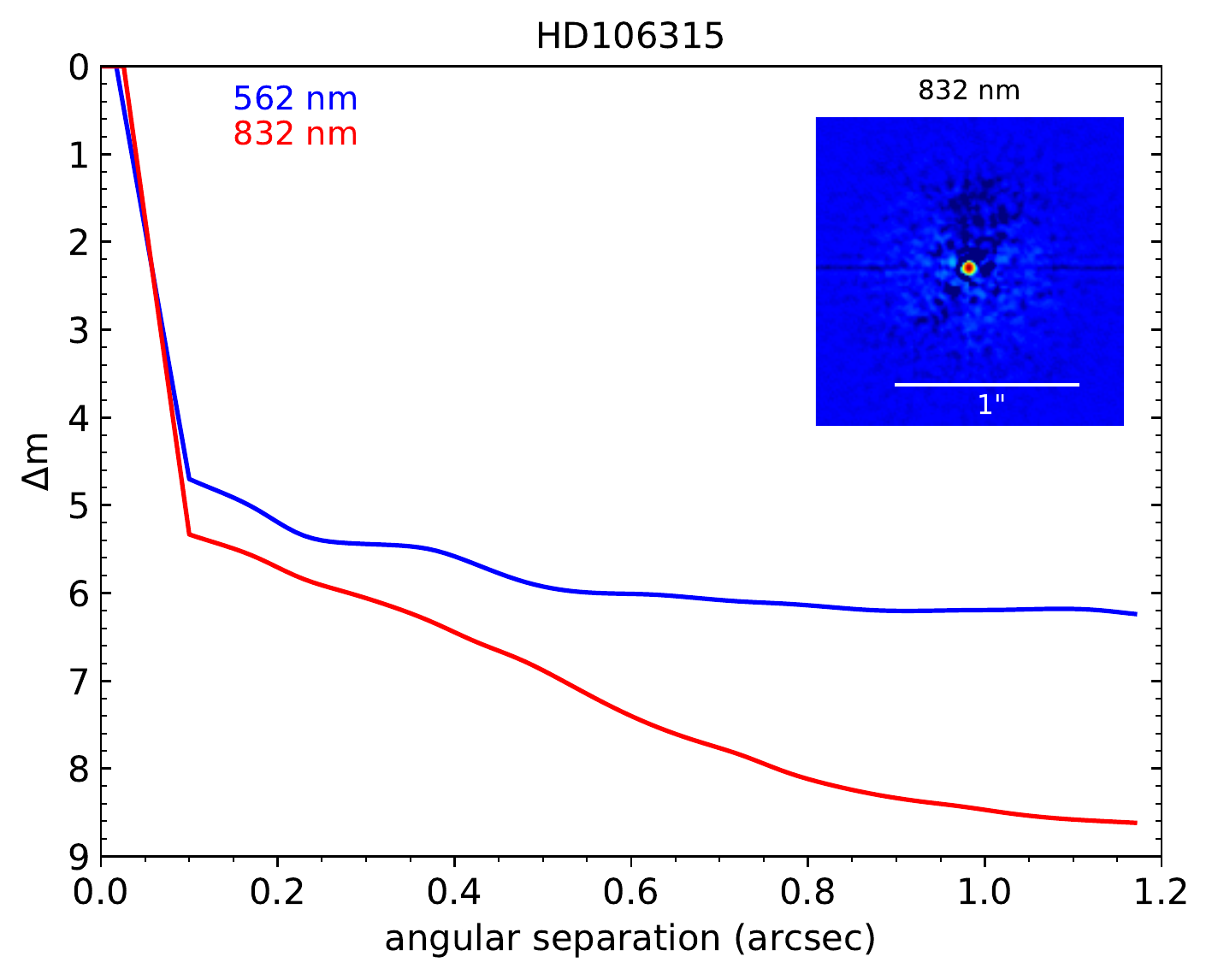}
\caption{\label{fig:hdimaging} 
Gemini-S/Zorro speckle-imaging contrast curve for \hd\ in 832nm (red) and 562nm (blue) including an inset image of the 832nm observation. No stellar companions or background sources are seen in these data.
}
\end{center}
\end{figure}

We observed HD 106315 on 2019 June 20 UT using the Zorro speckle interferometric instrument\footnote{https://www.gemini.edu/sciops/instruments/alopeke-zorro/} mounted on the 8-meter Gemini South telescope located on the summit of Cerro Pachon in Chile. Zorro simultaneously observes in two bands, one centered at 832nm with a width of 40nm and the other centered at 562nm with a width of 54nm, obtaining diffraction limited images with inner working angles 0.017 and 0.026 arcseconds, respectively. Our data set consisted of 3 minutes of total integration time taken as sets of 1000 $\times$ 0.06 sec images. All the images were combined and subjected to Fourier analysis leading to the production of final data products including speckle reconstructed imagery \citep[see][]{Howell2011}. Figure~\ref{fig:hdimaging} shows the 5-sigma contrast curves in both filters for the Zorro observation and includes an inset showing the 832 nm reconstructed image. The speckle imaging results confirm \hd\ to be a single star to contrast limits of 5--8.6 magnitudes, ruling out main sequence companions fainter than \hd\ itself within the spatial limits of 2 to 125 AU.

\section{Stellar Activity Analysis}
\label{sec:stellaractivity}

Variability in the brightness and velocity fields across the stellar disk results in line shape variations and apparent radial velocity shifts. Stellar activity with timescales comparable to planet orbital periods is a particular problem for radial velocity analyses as these signals can appear as additional Keplerian signals or can affect the fit amplitudes of the planet signals \citep[eg.][]{Fulton2015}. For our two systems, we focus on the component of stellar activity related to stellar rotation, as these signals have similar timescales to the transiting planet signals. 

Stellar activity can be tracked in radial velocity data using certain stellar lines as activity indicators. The Calcium II H\&K lines are often used for this purpose (S$_{\rm{HK}}$, \citealp{Isaacson2010}), whereas H-alpha may be more successful for cooler stars \citep{Robertson2013}. Another method is to use photometry to characterize the stellar activity and then subsequently fold the activity information into radial velocity fits \citep{Haywood2014}. For the Sun, there is a connection between stellar activity information derived from photometry, activity indicators, and radial velocity data \citep{Kosiarek2020}.
Here we investigate how stellar activity manifests in the \ktwo\ light curve, the Calcium II H\&K and H-alpha stellar lines, and our radial velocity data. 

\subsection{GJ 9827 Stellar Activity}

\begin{figure}[htbp]
\centering
\includegraphics[trim={0.5cm 0cm 2.5cm 0.5cm},clip,width=0.48\textwidth]{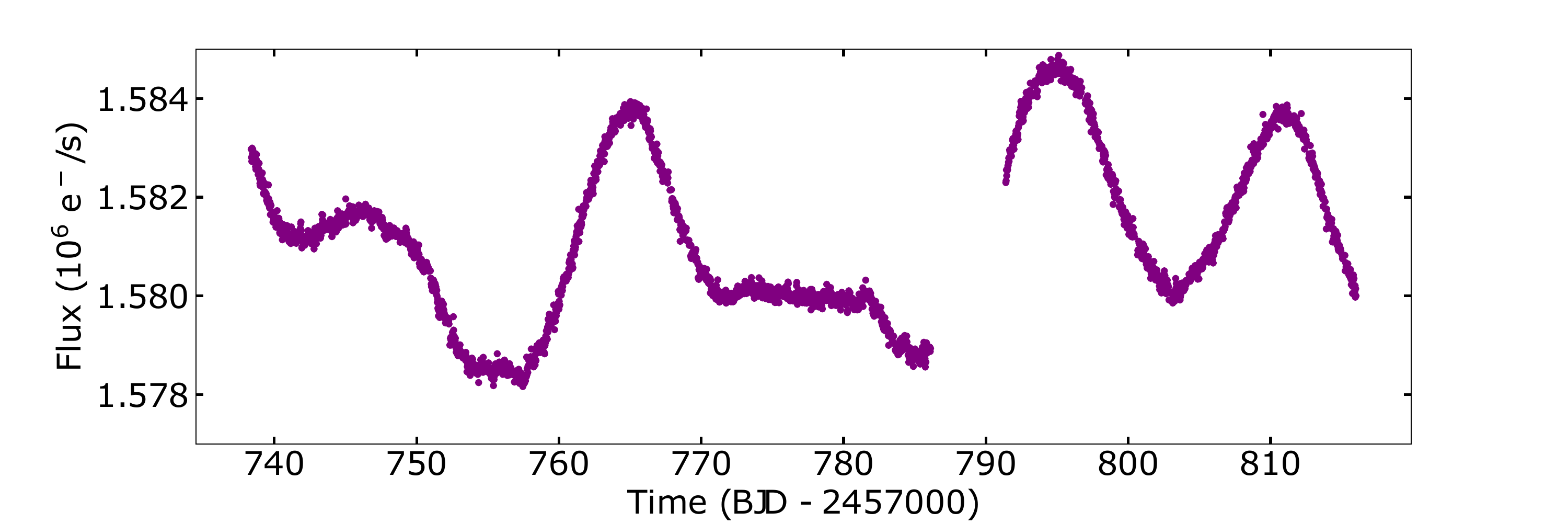}
\includegraphics[trim={0.5cm 3cm 2.5cm 4.5cm},clip,width=0.48\textwidth]{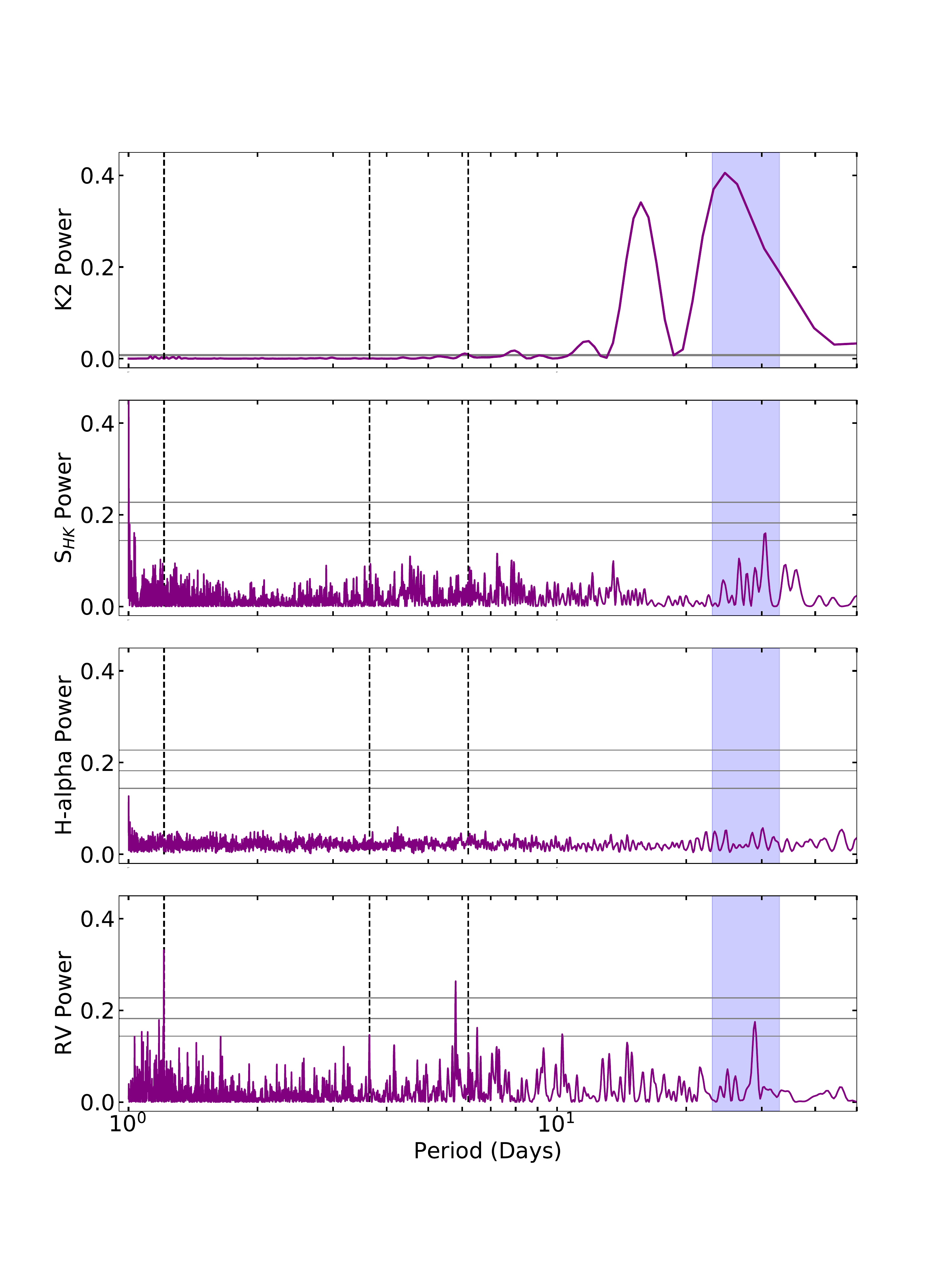}
\begin{center}
\caption{\label{fig:gjactivity} 
Activity analysis for \gj\ from K2 photometry and HIRES spectroscopy. There are clear stellar rotation and active region evolution signals visible by eye in the K2 photometry.
The Lomb-Scargle periodograms of the K2 photometry, S$_{\rm{HK}}$, H-alpha, and radial velocity data include false alarm probabilities of 0.5, 0.1, 0.01 (horizontal lines), stellar rotation (blue shaded area), and planet orbital periods (dashed lines). 
There is a stellar rotation signal at 30 days in the S$_{\rm HK}$ and radial velocity data, consistent with the broad peak in the K2 photometry. 
}
\end{center}
\end{figure}

The \ktwo\ light curve for \gj\ shows quasi periodic variation with signs of active region evolution between rotation cycles (Figure~\ref{fig:gjactivity}). The K2 photometry shown in this paper was produced using k2phot \citep{Petigura2015,Petigura2017}. A Lomb-Scargle periodogram of the \ktwo\ data shows two strong peaks around 15 and 30 days consistent with previous works, one peak is likely the rotation period and the other a harmonic. We consider both peaks since stellar rotation periods often do not appear as the highest peak in a periodogram \citep{Nava2019arxiv}. The shorter period is favored by \citet{Niraula2017} from the $v\sin i$ measurement, whereas the longer period is favored by \citet{Rodriguez2018,Teske2018,PrietoArranz2018, Rice2019} from a combination of periodogram, autocorrelation, and Gaussian process analyses on the light curve as well as from the inferred age of \gj.

The Keck/HIRES S$_{\rm HK}$ and radial velocity data shown in Figure~\ref{fig:gjactivity}, both reveal a tenuous stellar rotation signal at 30 days, consistent with the longer peak in the \ktwo\ light curve peridogram. In agreement with previous findings, we conclude that this 30 day signal is likely caused by stellar rotation, as it is present in both the S$_{\rm HK}$ data and the photometry. Since there is power at the same period in our radial velocity data, we need to account for this signal in our radial velocity analysis in order to derive accurate mass measurements for the planets. We mitigated this signal using a Gaussian process, as described below in Section~\ref{sec:gjgp}. 

\subsection{HD 106315 Stellar Activity}
Similar to \gj, we aim to understand the stellar activity component of the radial velocity data through investigating the possible relationships between the \ktwo\ light curve, the Calcium II H\&K and H-alpha stellar lines, and our radial velocity data. 
The projected rotational velocity measurement ($v \sin i = 13.2\pm1$ km~s$^{-1}$) combined with the obliquity measurement ($\lambda = -10.9\pm3.7$, \citealp{Zhou2018}) suggests a stellar rotation period of 4.78$\pm$0.15 days.

\begin{figure}[htbp]
\centering
\includegraphics[trim={0.5cm 27cm 2.5cm 1cm},clip,width=0.48\textwidth]{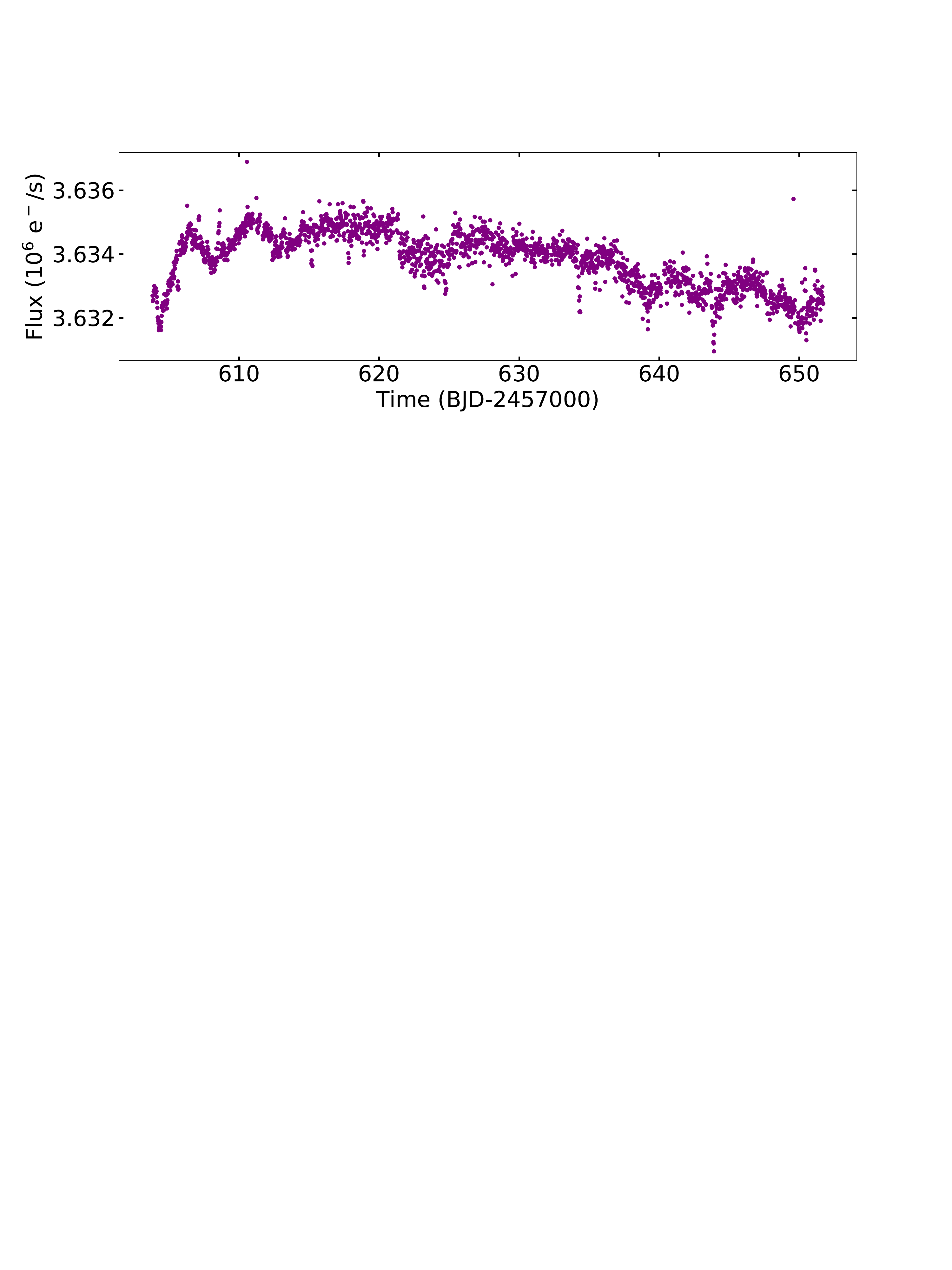}
\includegraphics[trim={0.5cm 3cm 2.5cm 4.5cm},clip,width=0.48\textwidth]{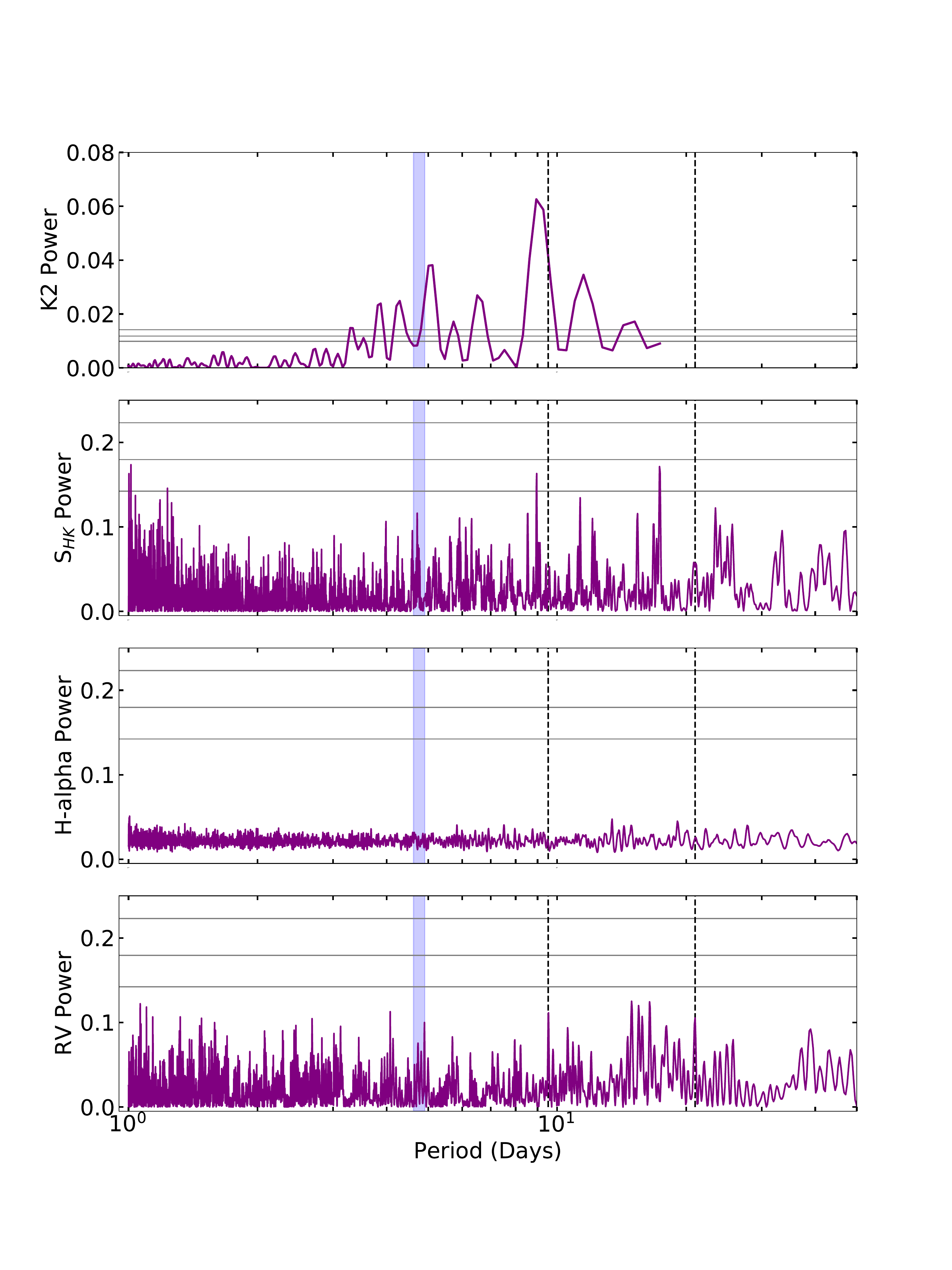}
\begin{center}
\caption{\label{fig:hdactivity} 
Activity analysis for \hd\ from K2 photometry and HIRES spectroscopy. 
The Lomb-Scargle periodograms of the photometry, S$_{\rm{HK}}$, H-alpha, and radial velocity data include false alarm probabilities of 0.5, 0.1, 0.01 (horizontal lines), stellar rotation period (thick blue line), and planet orbital periods (dashed lines). 
There are peaks near the rotation period and second harmonic in the K2 photometry, we find no similar peaks in the HIRES activity indicators or radial velocity data. 
}
\end{center}
\end{figure}

\hd\ was observed in \ktwo\ Campaign 10; this campaign had a 14 day data gap resulting in 49 days of contiguous data. With a 4.8 day rotation period, the shorter campaign should not impact our conclusions about stellar activity from this photometry.
The \ktwo\ light curve (Figure~\ref{fig:hdactivity}) has low photometric variability; the periodogram shows a small peak near the stellar rotation period at 4.8 days and a larger peak at the second harmonic of the rotation period at 9.6 days.

We next investigated the potential radial velocity signal from the stellar rotation by examining the S$_{\rm HK}$ and H-alpha data in the HIRES spectra (Table~\ref{tab:hdrvs}). We find no significant peaks near 4.8 days or elsewhere in Lomb-Scargle periodograms of the HIRES activity indicators and radial velocity data (Figure~\ref{fig:hdactivity}). 
The absence of these signals suggests that the stellar rotation is not contributing a significant stellar activity signal to the radial velocity measurements, potentially attributed to the low spot coverage of this F star \citep[$<1\%$,][]{Kreidberg2020}.  

\subsection{Ground-based Photometry}

Stellar photometry of both systems was collected from the Fairborn Observatory in Arizona to lengthen the photometry baseline from which to look for stellar variability. 

Photometry of \gj\ was collected with the Tennessee State University Celestron C14 0.36 m Automated Imaging Telescope \citep[AIT,][]{Henry1999,Eaton2003}.
A total of 74 observations were collected from 2018 September 22 to 2020 January 27th with the Cousins R filter (Table~\ref{tab:gjphot}). The differential magnitudes were computed by subtracting the average brightness of 7 comparison stars in the same field of view. A frequency spectrum of the observations show no significant periodicities between 1 and 100 days; the observations scatter about their mean with a standard deviation of 0.00372~mag. 

Photometry of \hd\ was collected with the T12 0.80~m Automatic Photoelectric Telescope (APT); the T12 APT is essentially identical in construction and operation to the T8 0.8~m APT described in \citet{Henry1999}. A total of 43 observations of \hd\ were collected between 2018 February 9 and 2018 June 7 in both the Stromgren b and y filters by T12's two-channel photometer (Table~\ref{tab:hdphot}). The two filters were averaged together into the (b+y)/2 ``filter" to increase the data precision. The differential magnitudes were calculated using three comparison stars: HD 105374, HD 105589, and HD 106965. A frequency spectrum of the observations show no significant periodicities between 1 and 100 days; the observations scatter about their mean with a standard deviation of 0.00256~mag. 

\section{Radial Velocity Analysis}
\label{sec:rv}

We analyzed the radial velocity data for these two systems with \texttt{radvel}\footnote{\url{https://radvel.readthedocs.io/}} \citep{Fulton2018}. \texttt{radvel} models Keplerian orbits and optional Gaussian processes to fit radial velocity data. The fit is performed through a maximum-likelihood function and errors are determined with an MCMC analysis. We use the default number of walkers, number of steps, and criteria for burn-in and convergence as described in \citet{Fulton2018}. 

For both systems, we first model the radial velocity data including circular Keplerian orbits for all of the transiting planets; we include a Gaussian prior on the orbital period ($P$) and time of transit ($T_{\rm conj}$) from our updated ephemerides in Section~\ref{sec:spitzer}. The semi-amplitudes ($K$) reported from these analyses refer to the motion of the star induced by the orbiting planet. Afterwards, we test models including a trend ($\dot{\gamma}$), curvature ($\ddot{\gamma}$), and planet eccentricities ($e$, $\omega$). We used the Akaike information criterion corrected for small samples sizes (AIC) to evaluate if the fit improved sufficiently to justify the additional free parameters; a lower AIC indicates an improved fit.

\subsection{Radial Velocity Analysis for \gj}
\label{sec:gjgp}

There is evidence of stellar activity in our radial velocity data from the periodogram analysis in Section~\ref{sec:stellaractivity}. 
We include a Gaussian process with a quasi-periodic kernel to model this activity signal in our radial velocity fit. The kernel has the form

\begin{equation}
\label{eq:kernel}
    k(t,t') = \eta_1^2 \ \rm{exp} \left[-\frac{(t-t')^2}{\eta_2^2}-\frac{sin^2(\frac{\pi(t-t')}{\eta_3})}{2 \eta_4^2}\right],
\end{equation}
where the hyperparameter $\eta_1$ is the amplitude of the covariance function, $\eta_2$ is the active region evolutionary time scale, $\eta_3$ is the period of the correlated signal, and $\eta_4$ is the length scale of the periodic component. 
We explore these hyperparameters for this system by performing a maximum likelihood fit to the \ktwo\ light curve, S$_{\rm HK}$, and H-alpha data with the quasi-periodic kernel (Equation~\ref{eq:kernel}), then determine the errors through an MCMC analysis. 

The \ktwo\ light curve fit is well constrained by the Gaussian process and produces a stellar rotation period consistent with the periodogram analysis of this data ($\eta_3$=28.62$^{+0.48}_{-0.38}$). 
The H-alpha data has very low variation; it is not well fit by this kernel and does not produce meaningful posteriors. 

The S$_{\rm HK}$ data is well fit by this quasi-periodic kernel and produces a stellar rotation period ($\eta_3$) consistent with our periodogram analysis in Section~\ref{sec:stellaractivity}. The photometry and the S$_{\rm HK}$ data both produce consistent posteriors; we choose to adopt the posteriors from the S$_{\rm HK}$ fit because these data are taken simultaneously with the radial velocity data and are therefore a direct indicator of the chromospheric magnetic activity. 
The posteriors on the parameters from our S$_{\rm HK}$ fit are: $\gamma_{S_{\rm HK}}$ = $0.646^{+0.027}_{-0.026}$, $\sigma_{S_{\rm HK}}$ = $0.0183^{+0.0035}_{-0.0032}$, $\eta_1$ = $0.079^{+0.017}_{-0.012}$, $\eta_2$ = $94^{+50}_{-25}$ days, $\eta_3$ = $29.86^{+0.78}_{-0.83}$ days, and $\eta_4$ = $0.587^{+0.14}_{-0.096}$.

We then performed a Gaussian process fit on the radial velocity data including priors on $\eta_2$, $\eta_3$, and $\eta_4$ equivalent to the $S_{\rm HK}$ fit posteriors. We tested fits including a trend, curvature, and planet eccentricities but reject all of these models due to their higher AIC values. 
These tested fits resulted in semi-amplitudes for all three planets consistent to 1$\sigma$ for planets b and d, and 2$\sigma$ for planet c with the circular 3-planet Gaussian process fit.

\begin{figure}[!h]
\centering
\includegraphics[width=0.5\textwidth,keepaspectratio]{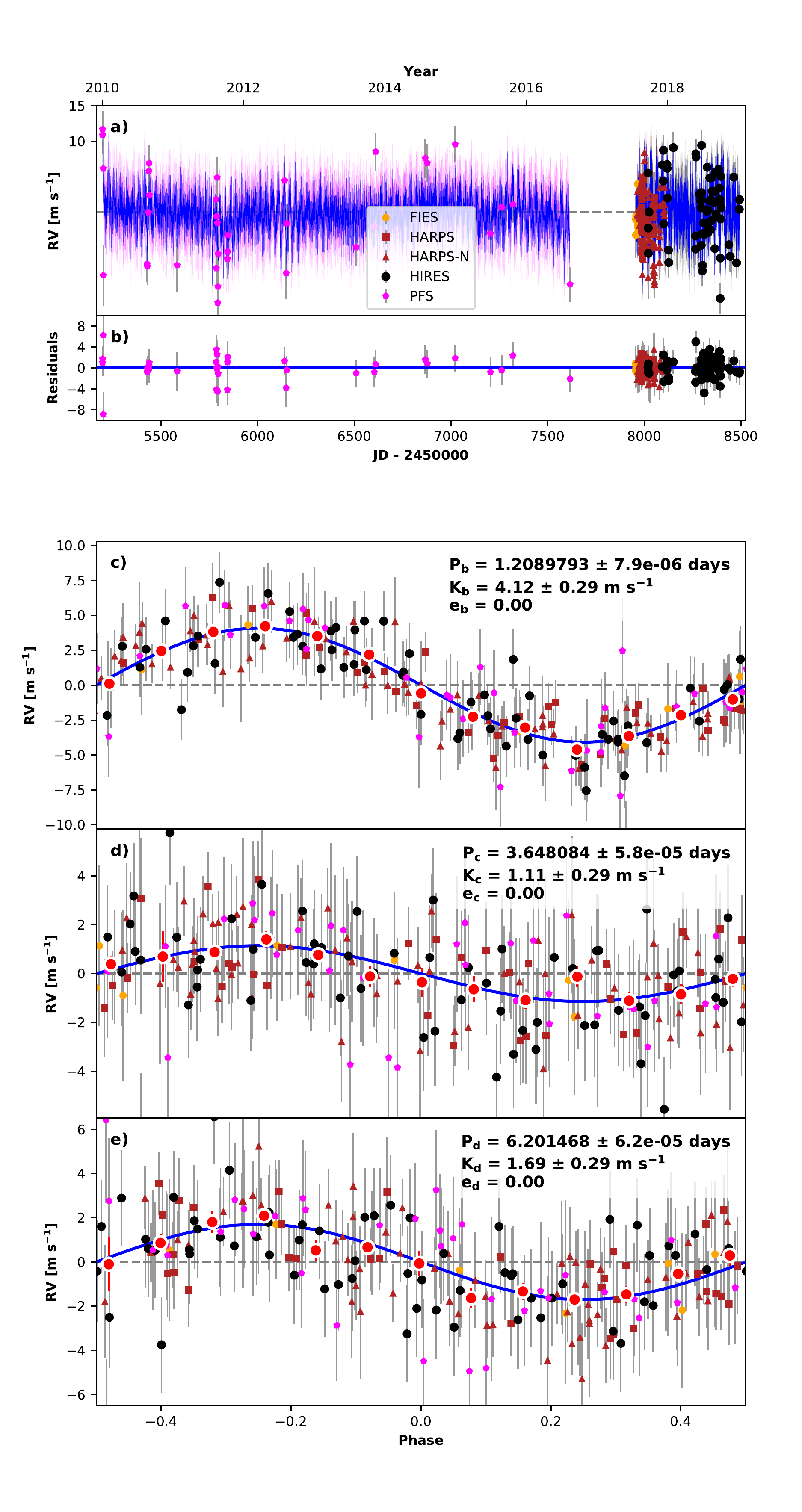}
\caption{Best-fit 3-planet Keplerian orbital model with a Gaussian process for GJ 9827. The thin blue line is the best-fit one-planet
model with the mean Gaussian process model; the colored area surrounding this line includes the 1$\sigma$ maximum-likelihood Gaussian process uncertainties. We add in quadrature the RV jitter terms listed in Table \ref{tab:gjparams} with the measurement uncertainties for all RVs.  {\bf b)} Residuals to the best fit 2-planet model. {\bf c)} RVs phase-folded to the ephemeris of planet b; the Keplerian orbit models for the other planets have been subtracted. Red circles are the same velocities binned in 0.08 units of orbital phase. {\bf d)} RVs phase-folded to the ephemeris of planet c. {\bf e)} RVs phase-folded to the ephemeris of planet d.
\label{fig:gjrv}
}
\end{figure}

We present our \gj\ results in Table~\ref{tab:gjparams}. We list the results from a circular 3-planet case with and without a Gaussian process for comparison, and adopt the fit including the Gaussian process shown in Figure~\ref{fig:gjrv}. We measure masses for these planets to be M$_b$=$4.87\pm 0.37$ \mearth, M$_c$=$1.92\pm 0.49$ \mearth, and M$_d$=$3.42\pm 0.62$ \mearth. 

\begin{deluxetable*}{lrrrr}
\tablecaption{GJ 9827 Radial Velocity Fit Parameters\label{tab:gjparams}}
\tablehead{\colhead{Parameter} & \colhead{Name (Units)} & \colhead{Keplerian fit} & \colhead{Gaussian Process fit (adopted)} }
\startdata
\sidehead{\bf{Orbital Parameters}}
$P_{b}$ & Period (days) & $1.2089765^{+2.2e-06}_{-2.3e-06}$ & $1.2089765\pm 2.3e-06$ \\
$T\rm{conj}_{b}$ & Time of Conjunction (BJD) & $2457738.82586\pm 0.00026$ & $2457738.82586\pm 0.00026$\\
R$_b$ & Radius (\rearth) & $\equiv$1.529$\pm$0.058 & $\equiv$1.529$\pm$0.058 \\
$e_{b}$ & Eccentricity & $\equiv0.0$ & $\equiv0.0$  \\
$\omega_{b}$ & Argument of Periapse & $\equiv0.0$ & $\equiv0.0$  \\
$K_{b}$ & Semi-Amplitude (m s$^{-1}$) &  $3.5\pm 0.32$ & $4.1\pm 0.3$\\
$a_b$  & Semimajor Axis (AU) &$0.01866\pm 0.00019$& $0.01866\pm 0.00019$\\
$M_b$  & Mass (M$_{\oplus}$) & $4.12^{+0.39}_{-0.38}$ & $4.87\pm 0.37$\\
$\rho_b$  & Density (g cm$^{-3}$) & $6.32^{+1.0}_{-0.87}$ & $7.47^{+1.1}_{-0.95}$\\
$P_{c}$ & Period (days) & $3.648095^{+2.5e-05}_{-2.4e-05}$ & $3.648095\pm 2.4e-05$ \\
$T\rm{conj}_{c}$ & Time of Conjunction (BJD) & $2457742.19927\pm 0.00071$ & $2457742.19929^{+0.00072}_{-0.00071}$ \\
R$_c$ & Radius (\rearth) & $\equiv$1.201$\pm$0.046 & $\equiv$1.201$\pm$0.046 \\
$e_{c}$ & Eccentricity & $\equiv0.0$ & $\equiv0.0$ \\
$\omega_{c}$ & Argument of Periapse & $\equiv0.0$  & $\equiv0.0$  \\
$K_{c}$ & Semi-Amplitude (m s$^{-1}$) & $1.28\pm 0.32$ & $1.13\pm 0.29$ \\
$a_c$  & Semimajor Axis (AU) & $0.03896^{+0.00039}_{-0.0004}$ & $0.03896^{+0.00039}_{-0.0004}$\\
$M_c$  & Mass (M$_{\oplus}$) & $2.17^{+0.54}_{-0.55}$ & $1.92\pm 0.49$\\
$\rho_c$  & Density (g cm$^{-3}$) & $6.9^{+2.0}_{-1.8}$ & $6.1^{+1.8}_{-1.6}$\\
$P_{d}$ & Period (days) & $6.20183\pm 1e-05$ & $6.20183\pm 1e-05$\\
$T\rm{conj}_{d}$ & Time of Conjunction (BJD) & $2457740.96114\pm 0.00044$ & $2457740.96114^{+0.00045}_{-0.00044}$ \\
R$_d$ & Radius (\rearth) & $\equiv$1.955$\pm$0.075 & $\equiv$1.955$\pm$0.075 \\
$e_{d}$ & Eccentricity & $\equiv0.0$ & $\equiv0.0$ \\
$\omega_{d}$ & Argument of Periapse & $\equiv0.0$  & $\equiv0.0$  \\
$K_{d}$ & Semi-Amplitude (m s$^{-1}$) & $1.63\pm 0.31$  & $1.7\pm 0.3$  \\
$a_d$  & Semimajor Axis (AU) & $0.0555^{+0.00056}_{-0.00057}$ & $0.0555^{+0.00055}_{-0.00057}$\\
$M_d$  & Mass (M$_{\oplus}$) & $3.29\pm 0.64$ & $3.42\pm 0.62$\\
$\rho_d$  & Density (g cm$^{-3}$) & $2.41^{+0.58}_{-0.52}$ & $2.51^{+0.57}_{-0.51}$\\
\hline
\sidehead{\bf{Instrument Parameters}}
$\gamma_{\rm HIRES}$ & Mean center-of-mass (m s$-1$) & $-1.87^{+0.38}_{-0.39}$ & $-2.4^{+1.3}_{-1.4}$ \\
$\gamma_{\rm HARPS}$ & Mean center-of-mass (m s$-1$) &  $31946.64\pm 0.37$ &  $31947.7^{+4.0}_{-3.6}$ \\
$\gamma_{\rm HARPS-N}$ & Mean center-of-mass (m s$-1$) & $31948.64^{+0.43}_{-0.42}$ & $31950.2^{+2.7}_{-2.6}$\\
$\gamma_{\rm PFS}$ & Mean center-of-mass (m s$-1$) & $0.28\pm 0.86$ & $0.6\pm 1.2$ \\
$\gamma_{\rm FIES}$ & Mean center-of-mass (m s$-1$) & $31775.5^{+1.1}_{-1.2}$ & $31775.6\pm 1.5$ \\
$\sigma_{\rm HIRES}$ & Jitter ($\rm m\ s^{-1}$) & $3.45^{+0.32}_{-0.27}$ & $2.15^{+0.49}_{-0.43}$\\
$\sigma_{\rm HARPS}$ & Jitter ($\rm m\ s^{-1}$) & $1.65^{+0.39}_{-0.35}$ & $0.91^{+0.44}_{-0.45}$\\
$\sigma_{\rm HARPS-N}$ & Jitter ($\rm m\ s^{-1}$) & $2.79^{+0.39}_{-0.35}$ &  $0.74^{+0.44}_{-0.45}$\\
$\sigma_{\rm PFS}$ & Jitter ($\rm m\ s^{-1}$) & $4.68^{+0.75}_{-0.62}$ & $4.0\pm 1.1$\\
$\sigma_{\rm FIES}$ & Jitter ($\rm m\ s^{-1}$) & $0.0001^{+0.0016}_{-0.0001}$ & $0.035^{+2.6}_{-0.035}$ \\
\hline
\sidehead{\bf{GP Parameters}}
$\eta_{1, \rm HIRES}$ & GP Amplitude ($\rm m\ s^{-1}$) & N/A & $3.7^{+1.2}_{-1.0}$ \\
$\eta_{1, \rm HARPS}$ & GP Amplitude ($\rm m\ s^{-1}$) & N/A & $5.3^{+3.5}_{-2.2}$ \\
$\eta_{1, \rm HARPS-N}$ & GP Amplitude ($\rm m\ s^{-1}$) & N/A & $5.1^{+2.3}_{-1.5}$\\
$\eta_{1, \rm PFS}$ & GP Amplitude ($\rm m\ s^{-1}$) & N/A & $4.0\pm 1.1$ &\\
$\eta_{1, \rm FIES}$ & GP Amplitude ($\rm m\ s^{-1}$) & N/A & $0.035^{+2.6}_{-0.035}$ \\
$\eta_{2}$ & Evolutionary Timescale (days) & N/A & $82^{+17}_{-14}$  \\
$\eta_{3}$ & Period of the Correlated Signal (days) & N/A & $28.62^{+0.48}_{-0.38}$  \\
$\eta_{4}$ & Lengthscale & N/A & $0.418^{+0.082}_{-0.065}$ \\
\enddata
\tablenotetext{}{Derived parameters use M$_*$=$0.593\pm0.018$, R$_*$=$0.579\pm0.019$ (This work), R$_{b}$/R$_*$=$0.02420\pm0.00044$, R$_{c}$/R$_*$=$0.01899\pm0.00036$, R$_{d}$/R$_*$=$0.03093\pm0.00062$ \citep{Rodriguez2018}. }
\end{deluxetable*}

\subsection{Radial Velocity Analysis for \hd}

\begin{figure}[!h]
\centering
\includegraphics[width=0.5\textwidth,keepaspectratio]{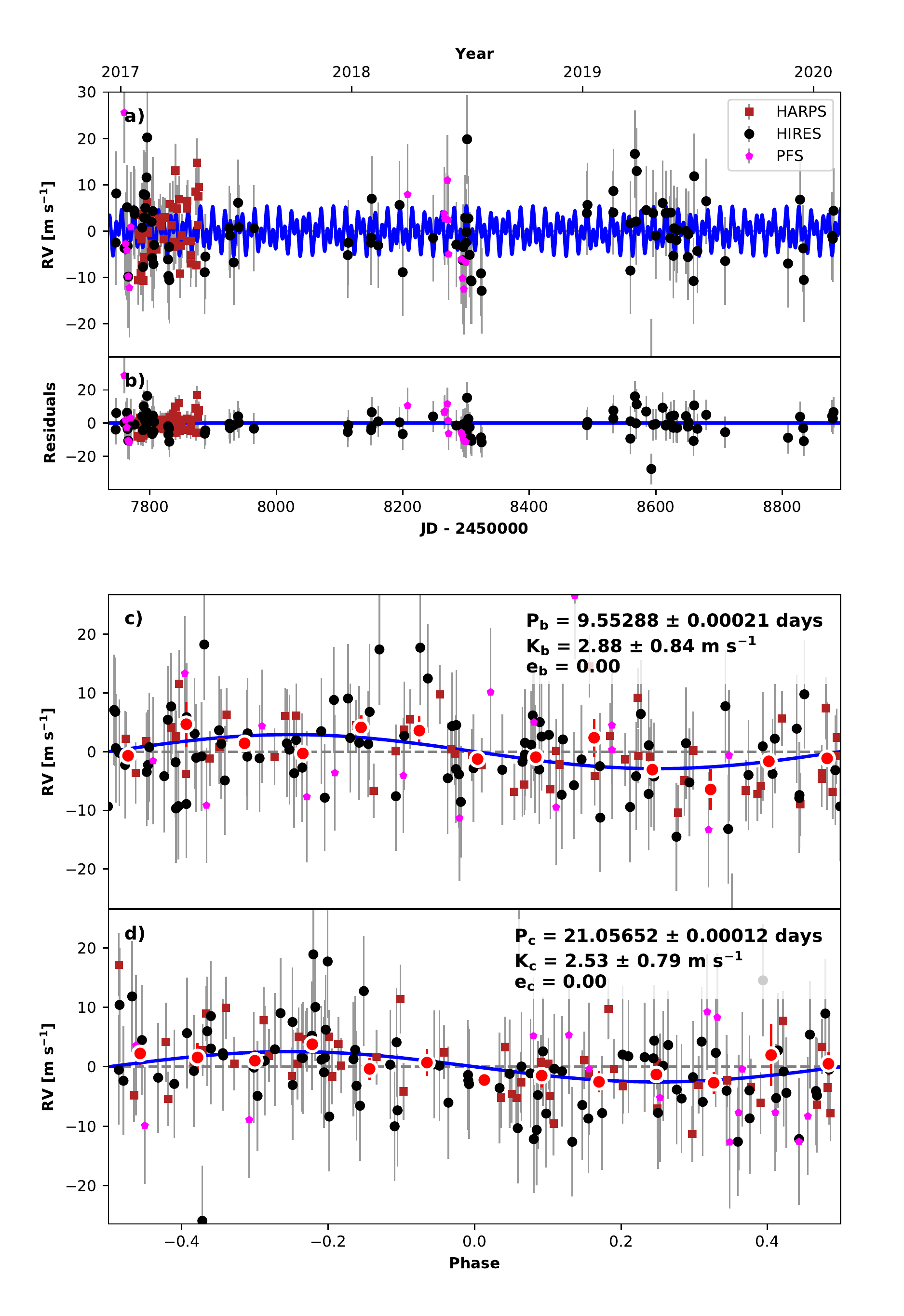}
\caption{Best-fit 2-planet Keplerian orbital model for HD 106315. The thin blue line is the best fit 2-planet model. We add in quadrature the RV jitter terms listed in Table \ref{tab:hdparams} with the measurement uncertainties for all RVs.  {\bf b)} Residuals to the best fit 2-planet model. {\bf c)} RVs phase-folded to the ephemeris of planet b with the orbit model of planet c subtracted. Red circles are the same velocities binned in 0.08 units of orbital phase. {\bf d)} RVs phase-folded to the ephemeris of planet c.
\label{fig:hdrv}
}
\end{figure}

For \hd, the circular 2-planet fit is favored by the AIC over fits with a trend, curvature, or planet eccentricities; results are listed in Table~\ref{tab:hdparams} and the fit is displayed in Figure~\ref{fig:hdrv}.
In agreement with \citet{Barros2017}, we do not see evidence of the trend suggested in \citet{Crossfield2017} with an AIC value 1.25 larger than the circular case. 
We determine masses for the HD 106315 system to be M$_b$=$10.5\pm 3.1$ \mearth\ and M$_c$=$12.0\pm3.8$ \mearth.

\begin{deluxetable*}{lrrr}
\tablecaption{HD 106315 Radial Velocity Fit Parameters\label{tab:hdparams}}
\tablehead{\colhead{Parameter} & \colhead{Name (Units)} & \colhead{Keplerian fit (adopted)} & \colhead{Gaussian Process fit}}
\startdata
\sidehead{\bf{Orbital Parameters}}
$P_{b}$ & Period (days) & $9.55288\pm 0.00021$ & $9.55288^{+0.00019}_{-0.00021}$  \\
$T\rm{conj}_{b}$ & Time of Conjunction (BJD) & $2457586.5476^{+0.0024}_{-0.0025}$  & $2457586.5479^{+0.003}_{-0.0026}$  \\
R$_b$ & Radius (\rearth) & $\equiv{2.40\pm0.20}$ & $\equiv{2.40\pm0.20}$ \\
$e_{b}$ & Eccentricity & $\equiv0.0$ & $\equiv0.0$ \\
$\omega_{b}$ & Argument of Periapse & $\equiv0.0$ & $\equiv0.0$ \\
$K_{b}$ & Semi-Amplitude (m s$^{-1}$) & $2.88^{+0.85}_{-0.84}$ & $2.91^{+0.79}_{-0.85}$\\
$a_b$  & Semimajor Axis (AU) & $0.0924^{+0.0011}_{-0.0012}$ &  $0.0924^{+0.0011}_{-0.0012}$ \\
$M_b$  & Mass (M$_{\oplus}$) & $10.5\pm 3.1$ & $10.6^{+2.9}_{-3.1}$\\
$\rho_b$  & Density (g cm$^{-3}$) & $4.1^{+1.9}_{-1.4}$  & $4.1^{+1.8}_{-1.4}$\\
$P_{c}$ & Period (days) & $21.05652\pm 0.00012$ & $21.05653\pm 0.00012$\\
$T\rm{conj}_{c}$ & Time of Conjunction (BJD) &  $2457569.01767^{+0.00097}_{-0.00096}$ &  $2457569.0178^{+0.0012}_{-0.001}$ \\
R$_c$ & Radius (\rearth) & $\equiv{4.379\pm0.086}$ & $\equiv{4.379\pm0.086}$ \\
$e_{c}$ & Eccentricity & $\equiv0.0$ & $\equiv0.0$\\
$\omega_{c}$ & Argument of Periapse & $\equiv0.0$ & $\equiv0.0$ \\
$K_{c}$ & Semi-Amplitude (m s$^{-1}$) & $2.53\pm 0.79$ & $2.61^{+0.74}_{-0.87}$\\
$a_c$  & Semimajor Axis (AU) & $0.1565^{+0.0019}_{-0.002}$  & $0.1565^{+0.0019}_{-0.002}$ \\
$M_c$  & Mass (M$_{\oplus}$) & $12.0\pm 3.8$ & $12.4^{+3.5}_{-4.2}$ \\
$\rho_c$  & Density (g cm$^{-3}$) & $0.78^{+0.26}_{-0.25}$ & $0.81^{+0.24}_{-0.27}$  \\
\hline
\sidehead{\bf{Instrument Parameters}}
$\gamma_{\rm HIRES}$ & Mean center-of-mass (m s$-1$) & $-2.48^{+0.96}_{-0.97}$ & $-2.7^{+1.0}_{-1.1}$   \\
$\gamma_{\rm HARPS}$ & Mean center-of-mass (m s$-1$) & $-3462.94^{+0.7}_{-0.71}$ &  $-3462.77^{+1.1}_{-0.87}$   \\
$\gamma_{\rm PFS}$ & Mean center-of-mass (m s$-1$) & $-2.9^{+2.8}_{-2.7}$ & $-2.5^{+3.2}_{-3.3}$\\
$\sigma_{\rm HIRES}$ & Jitter ($\rm m\ s^{-1}$) &  $8.33^{+0.85}_{-0.79}$ & $6.4^{+1.2}_{-1.1}$    \\
$\sigma_{\rm HARPS}$ & Jitter ($\rm m\ s^{-1}$) & $2.94^{+0.94}_{-1.0}$ &  $2.3^{+1.0}_{-1.4}$  \\
$\sigma_{\rm PFS}$ & Jitter ($\rm m\ s^{-1}$) & $9.4^{+2.6}_{-2.3}$  &  $4.0^{+4.6}_{-2.7}$ \\
\hline
\sidehead{\bf{GP Parameters}}
$\eta_{1, \rm HIRES}$ & GP Amplitude ($\rm m\ s^{-1}$) & N/A & $5.2^{+1.1}_{-1.7}$ \\
$\eta_{1, \rm HARPS}$ & GP Amplitude ($\rm m\ s^{-1}$) & N/A &  $2.3^{+1.0}_{-1.4}$\\
$\eta_{1, \rm PFS}$ & GP Amplitude ($\rm m\ s^{-1}$) & N/A &  $4.0^{+4.6}_{-2.7}$\\
$\eta_{2}$ & Evolutionary Timescale (days) & N/A &  $5.27^{+0.54}_{-0.65}$  \\
$\eta_{3}$ & Period of the Correlated Signal (days) & N/A &   $4.5^{+0.49}_{-0.65}$\\
$\eta_{4}$ & Lengthscale & N/A &  $0.56^{+0.036}_{-0.04}$ 
\enddata

\tablenotetext{}{Derived parameters use M$_*$=$1.154\pm0.043$, R$_*$=$1.269\pm0.024$ (This work), R$_{b}$/R$_*$=$0.01708\pm0.00135$ \citep{Crossfield2017}, R$_{c}$/R$_*$=$0.031636\pm0.0001834$ \citep{Kreidberg2020}. }
\end{deluxetable*}

In contrast with our \gj\ analysis, we choose not to include a Gaussian process in our HD 106315 fit as we do not see evidence for stellar rotation induced activity contamination in the activity indicators or radial velocity data. 
We suspect the low spot coverage of \hd\ \citep[$<1\%$,][]{Kreidberg2020} is why we see a small rotation signal in the photometry and a lack of this signal in our radial velocity data. 
\citet{Barros2017} does use a Gaussian process for their analysis of \hd. The derived Gaussian process period is 2.8 days and their full width half maximum (FWHM) measurements also show a similar periodicity leading them to believe that this signal arises from stellar activity. At the time, \citet{Zhou2018} had not yet measured the obliquity; therefore, \citet{Barros2017} hypothesized that this 2.8 day signal was the stellar rotation period or half of the rotation period.

If this signal is associated with stellar activity, it is possible that their high cadence radial velocity run is more sensitive to this activity than our data collection spanning multiple years. The HARPS measurements were collected on 47 nights over three months, whereas we have 94 nights of HIRES measurements over three years. 
It is also possible that the Gaussian process used by \citet{Barros2017} had fit spurious noise instead of a stellar activity signal; the 2.8 day signal is too short to be the rotation period or half of the rotation period. 
Hotter stars (T$_{\rm eff}>$6200 K) often have shallow convective envelopes and inefficient magnetic dynamos which result in fewer spots on the stellar surface \citep{Kraft1967}. Therefore, hotter stars like \hd\ may not have enough starspots for this type of Gaussian process to be effective.  

For completeness, we perform a Gaussian process fit on the \hd\ radial velocity data. We first fit the \ktwo\ data using a Gaussian process as this dataset showed periodicity at the stellar rotation period; the posteriors of this fit are: $\gamma_{K2}$ = $3633710^{+190}_{-200}$ e$^{-}$s$^{-1}$, $\sigma_{K2}$ = $117^{+16}_{-15}$ e$^{-}$s$^{-1}$, $\eta_1$ = $655^{+84}_{-68}$ e$^{-}$s$^{-1}$, $\eta_2$ = $5.17^{+0.66}_{-0.64}$ days, $\eta_3$ = $4.49^{+0.61}_{-0.26}$ days, $\eta_4$ = $0.55^{+0.04}_{-0.044}$. 
We then performed a Gaussian process fit on the radial velocity data including priors on $\eta_2$, $\eta_3$, and $\eta_4$ from the \ktwo\ fit posteriors. This fit results in semi-amplitudes consistent to 1$\sigma$ for both planets: the full results are shown in Table~\ref{tab:hdparams}. The Gaussian process fit has a higher AIC value ($\Delta$AIC=7.38) suggesting that Gaussian process parameters do not significantly improve the fit. For this reason, and as we do not see signs of stellar activity in our activity indicators or radial velocity data, we adopt the fit without a Gaussian process.

\subsection{Eccentricity Constraints}
\label{sec:ecc}

We explored the range of planet eccentricities consistent with system stability through N-body simulations as including eccentricity was not warranted in our radial velocity fits for either system. 
The literature papers on \gj\ assumed circular orbits for their fits \citep{PrietoArranz2018,Rice2019}. For \hd, \citet{Barros2017} includes eccentricity terms in their radial velocity analysis resulting in $e_b$=0.1$\pm$0.1 and $e_c$=0.22$\pm$0.15, although they do not discuss if including the eccentricity terms improve the fit. Our eccentric radial velocity fit for \hd\ resulted in $e_b$=0.18$\pm$0.17 and $e_c$=0.21$\pm$0.24,  consistent with \citet{Barros2017}. Though our eccentric fit had a higher AIC than the circular fit ($\Delta$AIC=6.22) suggesting that including eccentricity did not sufficiently improve the fit to justify the additional parameters.

We evaluated the stability of both systems using \texttt{spock} \citep{Tamayo2020}. \texttt{spock} predicts whether a given orbital configuration is stable by using \texttt{rebound} \citep{Rein2011} to simulate the first 10$^4$ orbits of a system and then calculating the probability that this system is stable for a full 10$^9$ orbits by comparing it to a wide sample of full simulations. These full simulations include mean-motion resonance, mutually inclined, and eccentric systems; the parameters are drawn from those typically encountered in current multiplanet systems.

We initialized both systems at the maximum likelihood values for the planet masses, orbital periods, times of conjunction, and stellar masses derived in this paper.

\begin{figure}[htb]
\centering
\includegraphics[trim={2cm 1cm 2cm 2cm},clip,width=0.5\textwidth]{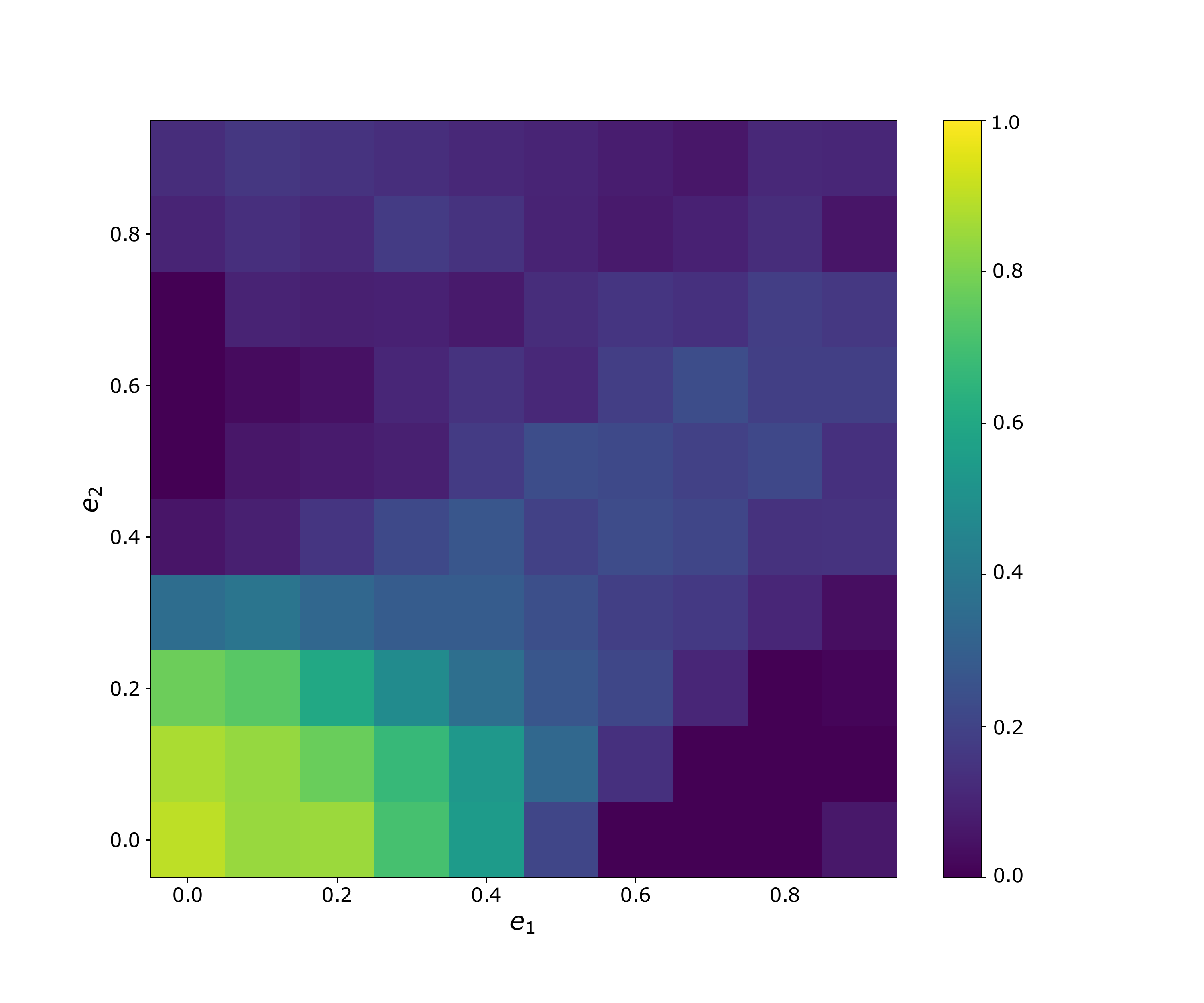}
\caption{Probability of stability for the \hd\ system. We examined the effect of planet eccentricity on the system's stability using \texttt{spock}. For each pair of eccentricities, we vary $\omega_1$ and $\omega_2$ from 0 to $2\pi$. The color of the box displays the average probability of stability across all $\omega$. 
\label{fig:hdecc}
}
\end{figure}

We then varied $e$ and $\omega$ for all planets to explore the stability of the system. 
For \hd, we varied $e_1$ and $e_2$ from 0.0 to 0.9 in steps of 0.1. At each eccentricity pair, we performed a grid of simulations varying $\omega_1$ and $\omega_2$ from 0 to 2$\pi$ in steps of $\frac{\pi}{5}$, resulting in 10,000 simulations. We then averaged over the simulated $\omega$ grid to calculate the average probability that a given eccentricity pair is stable (Figure~\ref{fig:hdecc}). 

\hdb\ and \hdc\ are in relatively close orbits at periods of 9.55 and 21.06 days; their orbits are unstable if either planet has a large eccentricity. The system has a probability of stability greater than 50\% when $e_1 \leq$ 0.4 and $e_2 \leq$ 0.3; the highest probability of stability is when both planets are in circular orbits. 

\begin{figure*}[htb]
\centering
\includegraphics[trim={4.6cm 0cm 5cm 0cm},clip,width=1.15\textwidth]{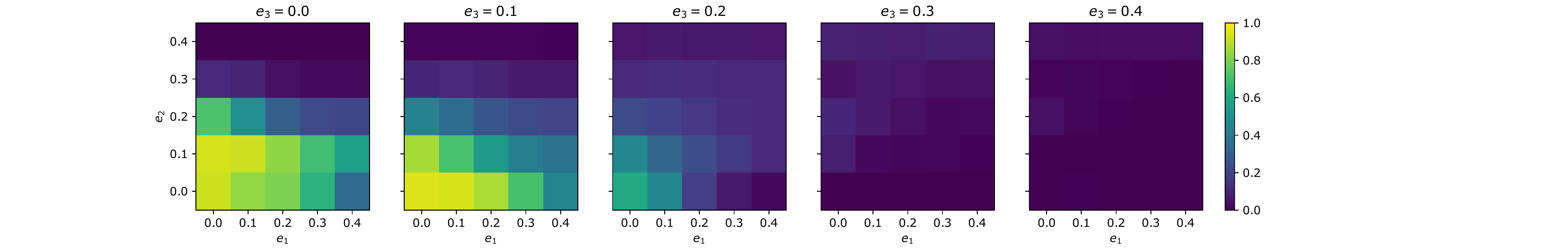}
\caption{Probability of stability for the \gj\ system. We examined the effect of planet eccentricity on the system's stability using \texttt{spock}. For each triplet of eccentricities, we vary $\omega_1$, $\omega_2$, and $\omega_3$ from 0 to $2\pi$. The color of the box displays the average probability of stability across all $\omega$. 
\label{fig:gjecc}
}
\end{figure*}

\gjb, \gjc, and \gjd\ are in even more closely-packed orbits at orbital periods of 1.2, 3.6, and 6.2 days. Therefore, for \gj, we varied $e_1$, $e_2$, and $e_3$ from 0.0 to 0.4 in steps of 0.1 as larger eccentricities for any of the three planets resulted in unstable orbits. 
At each eccentricity triplet we perform a grid of simulations varying $\omega_1$, $\omega_2$, and $\omega_3$ from 0 to 2$\pi$ in steps of $\frac{\pi}{5}$, creating a total of 125,000 simulations. We then averaged over the $\omega$ grid to calculate the average probability that a given eccentricity triplet is stable (Figure~\ref{fig:gjecc}). 

We find that the \gj\ system is unstable if $e_3\geq$ 0.3 and the system has very low stability at $e_3=$ 0.2. For $e_3\leq$ 0.1, the system can be stable with $e_2\leq$ 0.2 and $e_1\leq$ 0.4. This system has a smaller range of stable eccentricity values since the planets are packed closer together.

We then convert these eccentricity constraints to secondary eclipse timing constraints \citep[][equation 33]{Winn2010}. We note that the planets in these two systems are not particularly favorable targets for thermal emission spectroscopy based on the emission spectroscopy metric \citep[ESM,][\hdb: 3, \hdc: 6, \gjb: 14, \gjc: 4, \gjd: 6]{Kempton2018}.

From our eccentricity constraints and assuming $\omega$=0, the maximum offsets of the secondary eclipse time for \hdb\ and \hdc\ are 2.4 days and 4.0 days respectively. 
The maximum secondary eclipse timing offsets for the \gj\ system are 0.31 days, 0.46 days, and 0.39 days for planet b, c, and d respectively.

\section{Interior Bulk Compositions}
\label{sec:interior}

\begin{figure}[htb]
\centering
\includegraphics[trim={2cm 0cm 0cm 0cm},clip,width=0.5\textwidth]{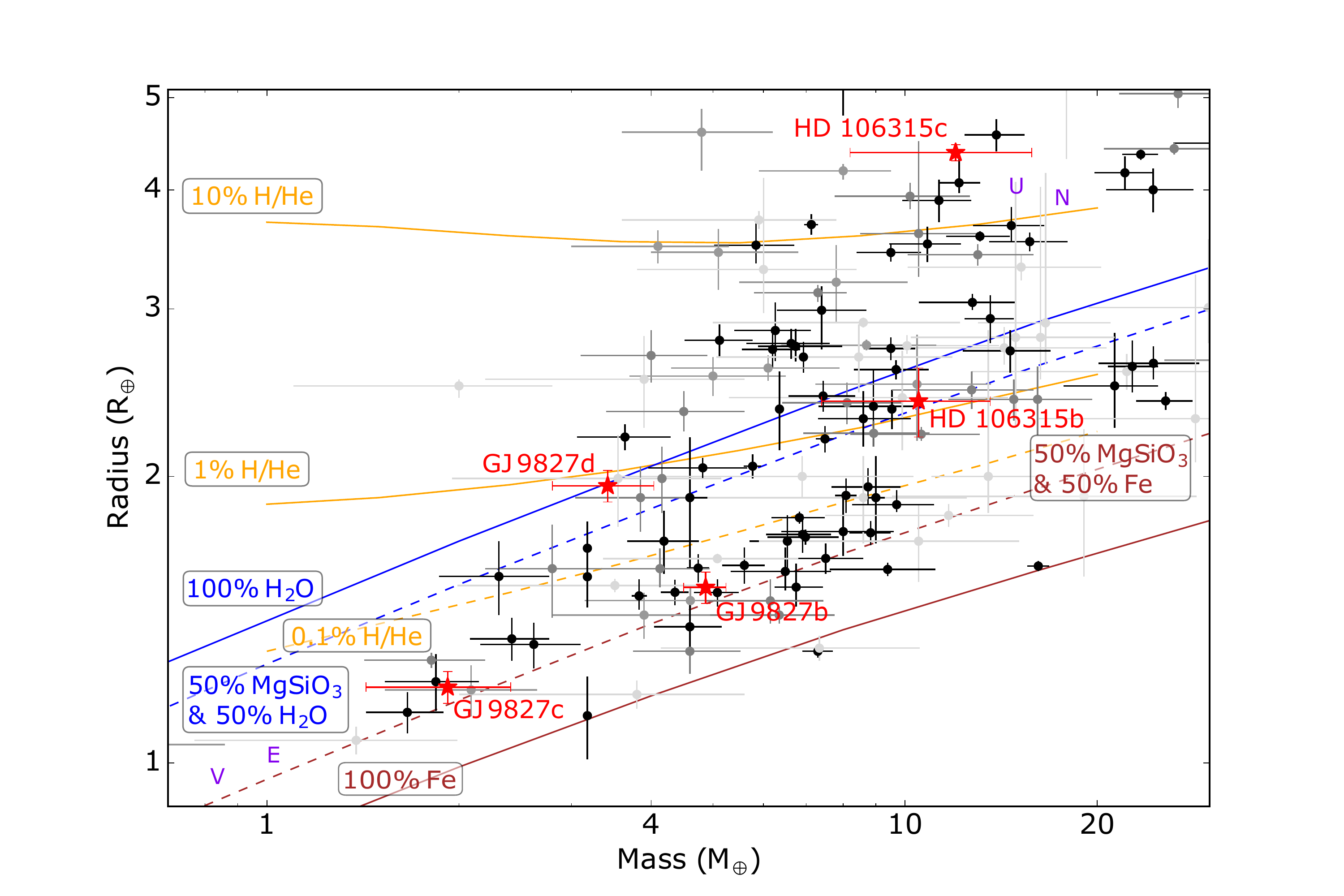}
\caption{Mass-radius diagram for planets between the size
of Earth and Neptune with greater than 2$\sigma$ measurements
(darker points for lower error). The lines show models of
different compositions \citep{Lopez2014,Zeng2016}. 
Our five planets are shown as red stars with 1$\sigma$ uncertainties. 
\label{fig:mrdiagram}
}
\end{figure}

To explore the interior compositions of these planets, we first visually compare their masses and radii to other known exoplanets on a mass-radius diagram (Figure~\ref{fig:mrdiagram}). \gjb\ ($\rho_{b}$=$7.5$ g cm$^{-3}$) and \gjc\ ($\rho_{c}$=$6.1$ g cm$^{-3}$) are both consistent with a 50/50 mixture of rock and iron.
\gjd\ ($\rho_{d}$=$2.5$ g cm$^{-3}$) and \hdb\ ($\rho_{b}$=$4.1$ g cm$^{-3}$) are consistent with either 100\% water or a rocky core with a 1\% H/He envelope. 
Lastly, \hdc\ ($\rho_{c}$=$0.8$ g cm$^{-3}$) is located near our solar system ice giant planets. It has a much lower density than \hdb, too low to be explained by water alone, and is consistent with having a $>$10\% H/He envelope. 

To further investigate the interior compositions of these planets,
we compared their masses and radii with model composition grids from \citet{Zeng2013,Lopez2014,Zeng2016}. 
We focus on two main compositions: Earth-like rock \& iron cores surrounded by H/He envelopes and mixtures of water, rock, and iron. Our results are tabulated in Table~\ref{tab:comp}.

\begin{deluxetable}{lcc}[b]
\tablecaption{Hydrogen/Helium and Water Mass Fraction Surrounding an Earth-like Core \label{tab:comp}}
\tablehead{\colhead{Planet} & \colhead{$f_{HHe}$ (\%)} & \colhead{$f_{H_2O}$ (\%)} }
\startdata
GJ 9827b & $0.02^{+0.01}_{-0.01}$ & $2.20^{+3.84}_{-1.69}$ \\
GJ 9827c & $0.01^{+0.01}_{-0.00}$ & $13.57^{+25.18}_{-10.40}$ \\
GJ 9827d & $0.54^{+0.20}_{-0.17}$ & $79.10^{+14.35}_{-20.14}$ \\ 
HD 106315b & $0.96^{+0.72}_{-0.51}$ & $54.29^{+29.06}_{-30.09}$ \\
HD 106315c & $12.74^{+1.11}_{-1.06}$ & $99.27^{+0.57}_{-1.25}$
\enddata
\end{deluxetable}

To calculate potential H/He mass fractions, we use the grids of thermal evolution models provided by Lopez \& Fortney (2014) which calculate the radius of a planet given varying incident fluxes relative to Earth ($S_{inc}/S_\oplus$), masses ($M_p/M_\oplus$), ages, and fractions $f_{HHe}$ of their masses contained in H/He envelopes surrounding Earth-like rock and iron cores. 
We use the \texttt{smint} (Structure Model INTerpolator) interpolation and envelope mass fraction fitting package, which we made publicly-available on GitHub\footnote{\url{https://github.com/cpiaulet/smint}}, in order to  solve the inverse problem of inferring a planet's envelope mass fraction from its incident flux, mass, age, and radius.

\begin{figure*}[htbp]
\centering
\begin{subfigure}[\gjd]{
\includegraphics[width=0.50\textwidth]{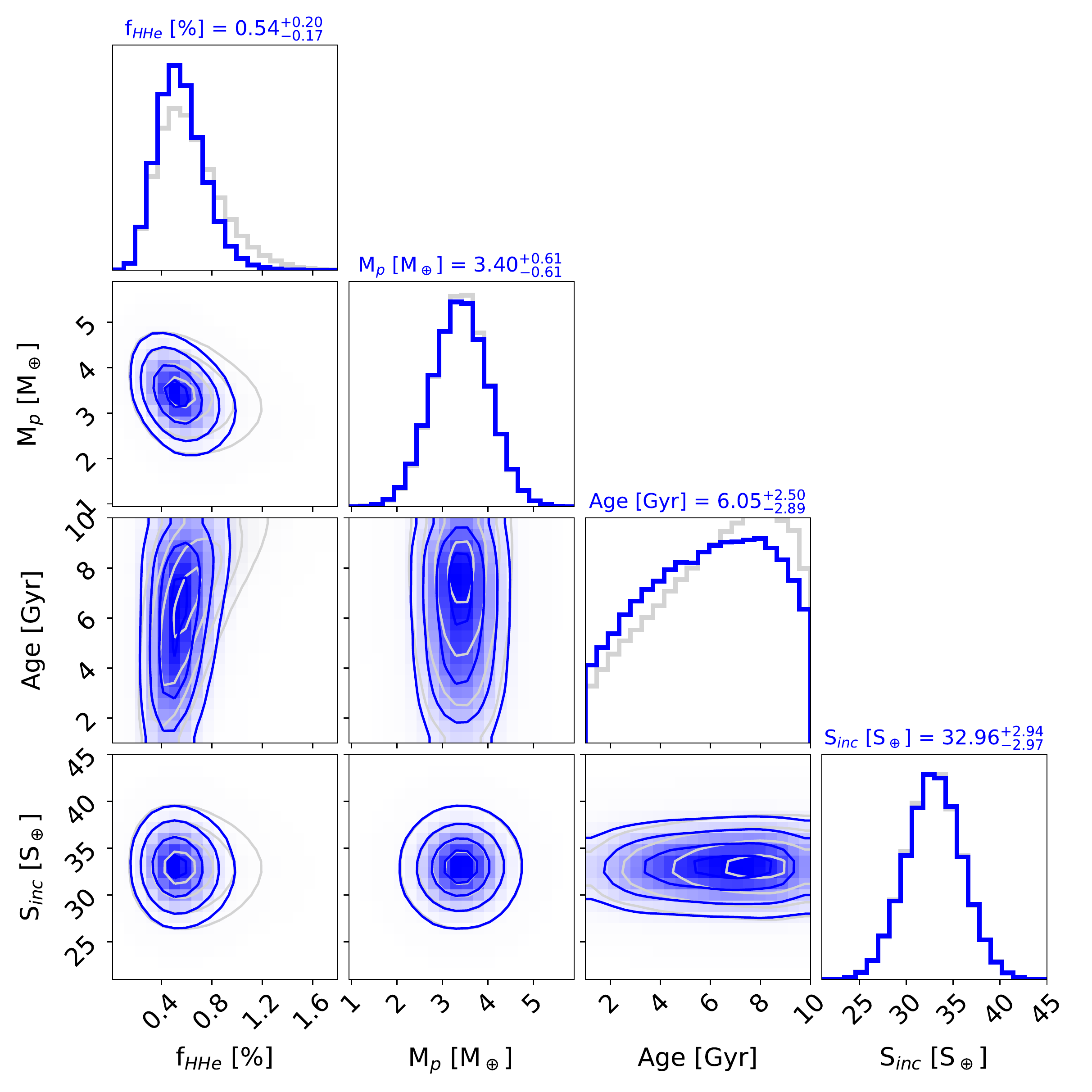}}
\end{subfigure}
\hspace{2cm}
\begin{subfigure}[\hdb]{
\includegraphics[width=0.48\textwidth]{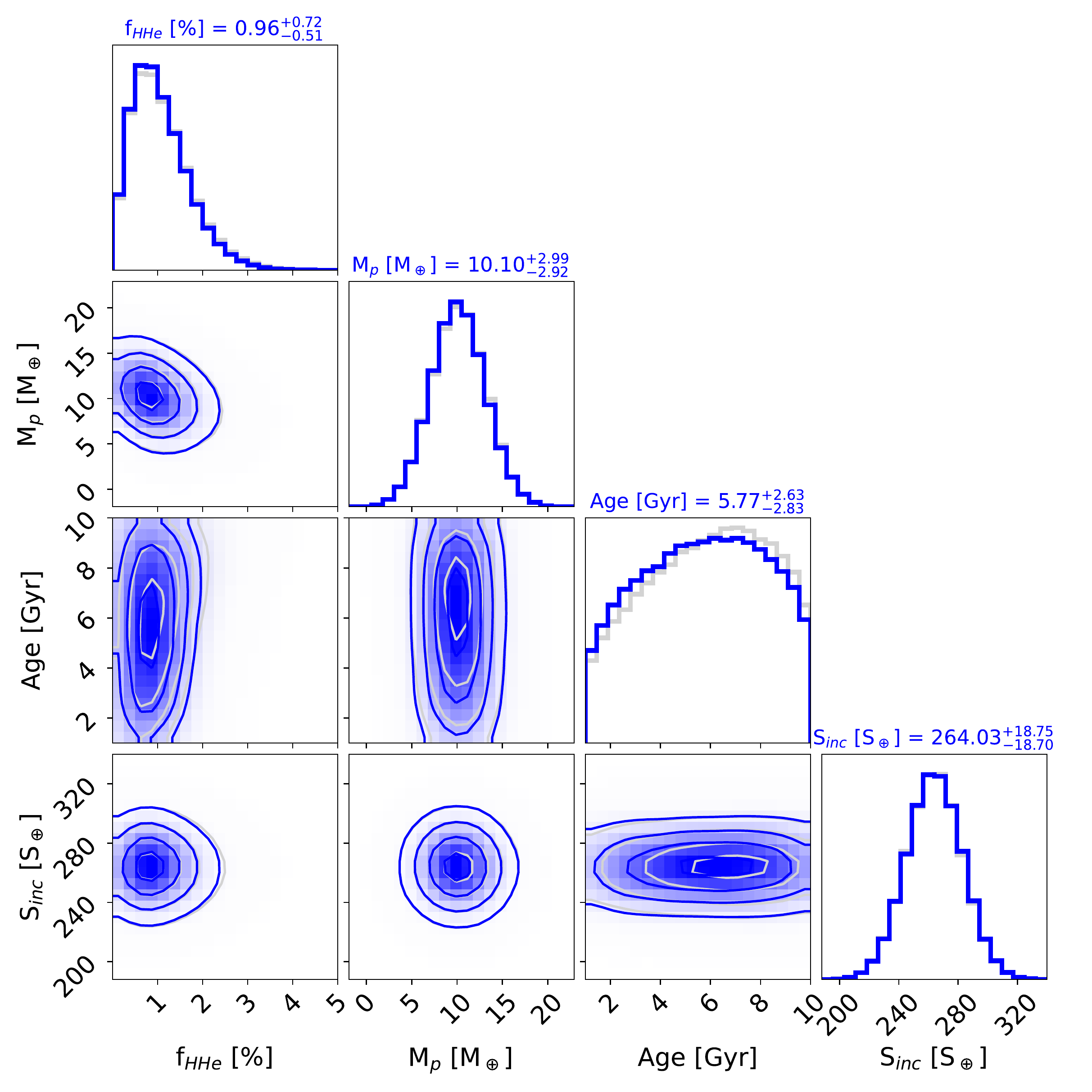}}
\end{subfigure}
\begin{subfigure}[\hdc]{
\includegraphics[width=0.48\textwidth]{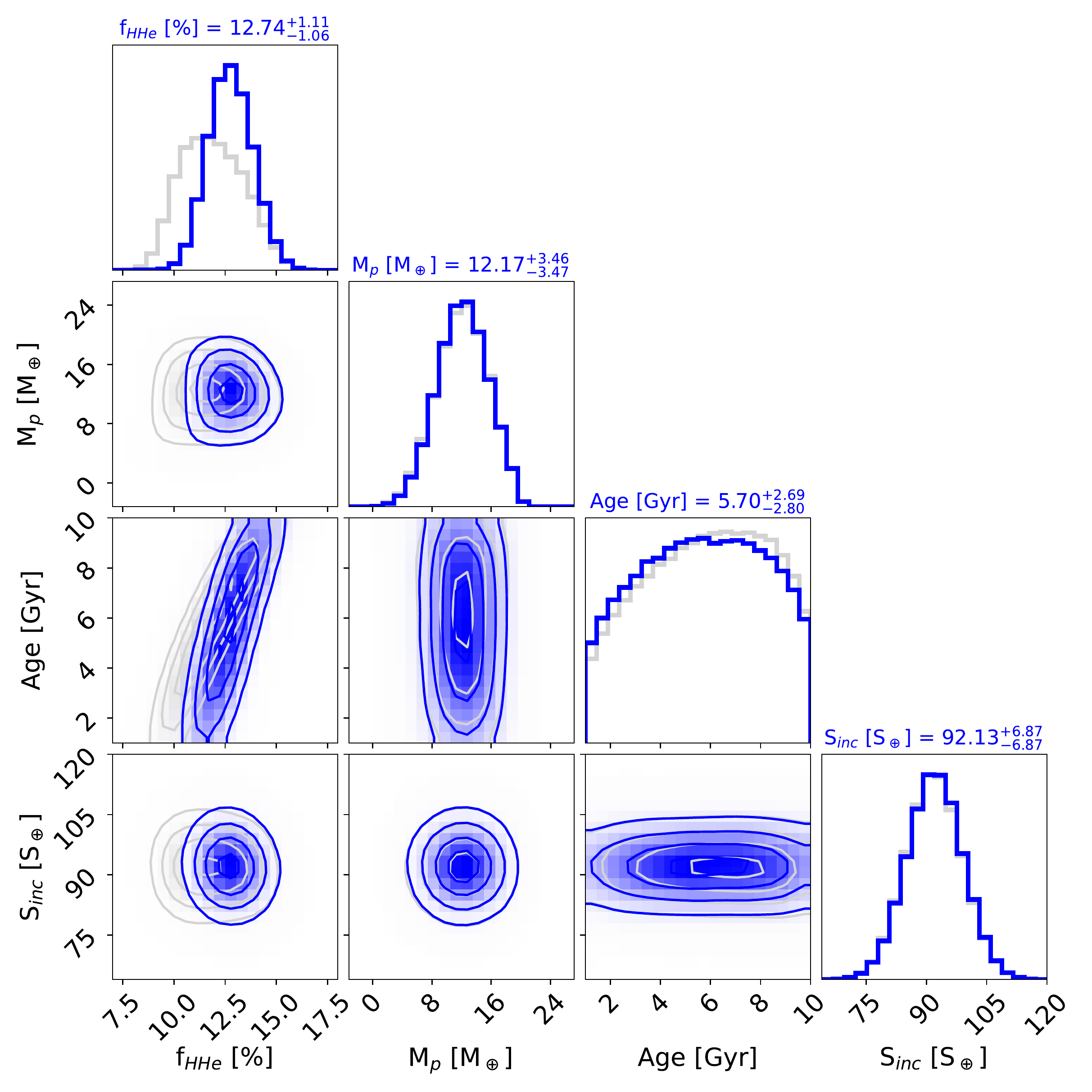}}
\end{subfigure}
\caption{\label{fig:hhe} Joint and marginalized posterior distributions on the fitted parameters for a 1$\times$ (50$\times$) solar metallicity H$_2$/He envelope atop an Earth-like core are shown in blue (gray). The median and $\pm 1\sigma$ constraints on the parameters for the 1$\times$ solar metallicity case are quoted above each marginalized distribution.}
\end{figure*}

\texttt{smint} performs linear interpolation over a grid of $f_{HHe}$, $\log_{10} M_p/M_\oplus$, system age and $\log_{10} S_{inc}/S_\oplus$ and returns the corresponding planet radius. We then run a MCMC that fits for the combination of $S_{inc}$, $M_p$, age, and $f_{HHe}$ that best matches the observed planet radius. 
We adopt Gaussian priors on $S_{inc}$ and $M_p$ informed by the stellar and planetary parameters. We use a uniform prior on the planet's envelope mass fraction over the range spanned by the \citet{Lopez2014} grids (from 0.1 to 20\%) and adopt a uniform prior on the system age from 1 to 10 Gyr. 
Each of the 100 chains is run for at least 10,000 steps, 60\% of which are discarded as burn-in. We make sure that, in each case, the chains have run for at least 50 times the maximum autocorrelation time recorded across all parameters and thus secure that our chains are converged and well sample the posterior PDFs. We display a corner plot for the three planets consistent with moderate H/He envelopes, \gjd, \hdb, and \hdc\ (Figure~\ref{fig:hhe}). \gjd\ and \hdb\ are both consistent with ~1\% H/He envelopes and \hdc\ is consistent with a 13\% H/He envelope. 

To fit for the water mass fractions ($f_{H_2O}$), we use the implementation of the \citet{Zeng2016} two-component (water+rock) model grid in \texttt{smint} (Table~\ref{tab:comp}). The MCMC process is analog to that used to fit for $f_{HHe}$, adopting a uniform prior on the water mass fraction (0--100 \%) and a Gaussian prior on the planet mass. We match the observed planet radius via a Gaussian likelihood.

\begin{figure*}[htbp]
\centering
\begin{subfigure}[\gjb]{
\includegraphics[width=0.45\textwidth]{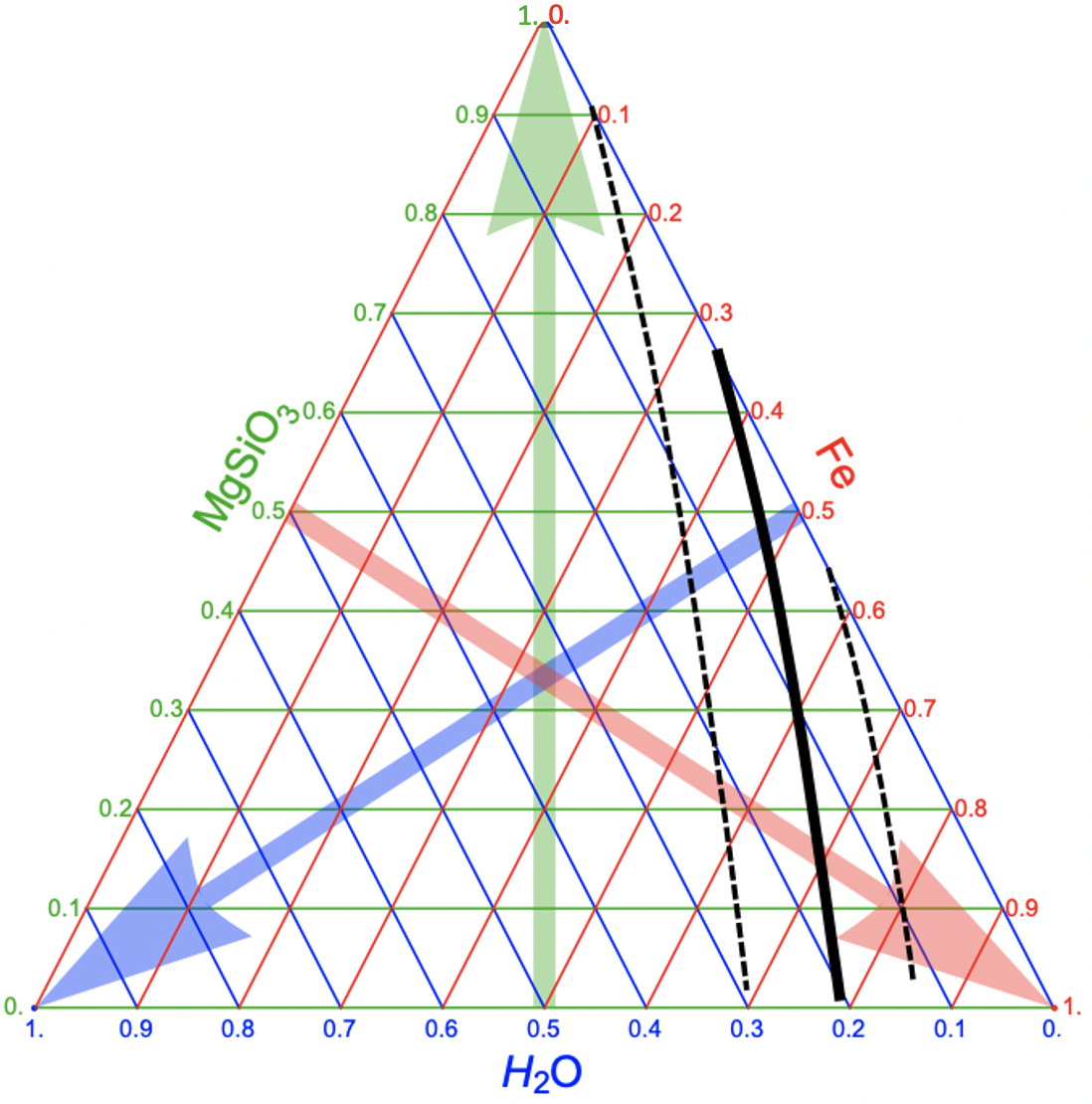}}
\end{subfigure}
\begin{subfigure}[\gjc]{
\includegraphics[width=0.45\textwidth]{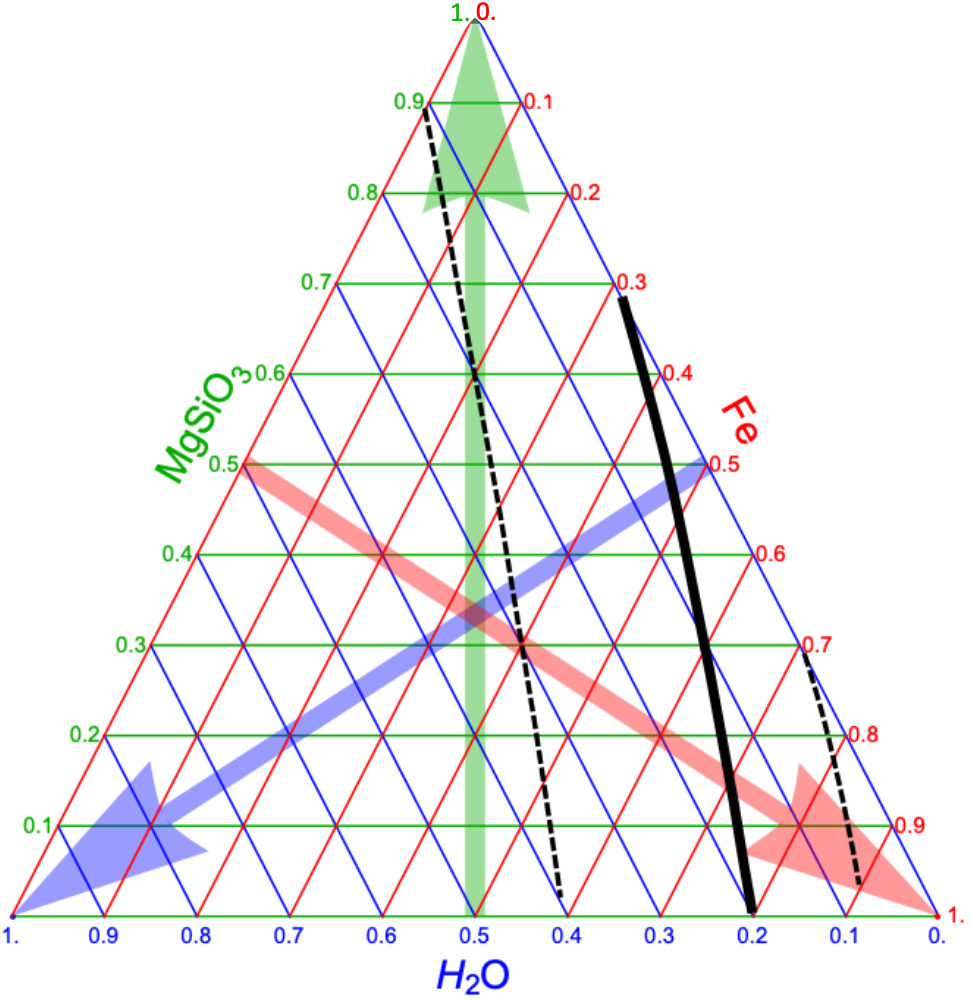}}
\end{subfigure}
\begin{subfigure}[\gjd]{
\includegraphics[width=0.45\textwidth]{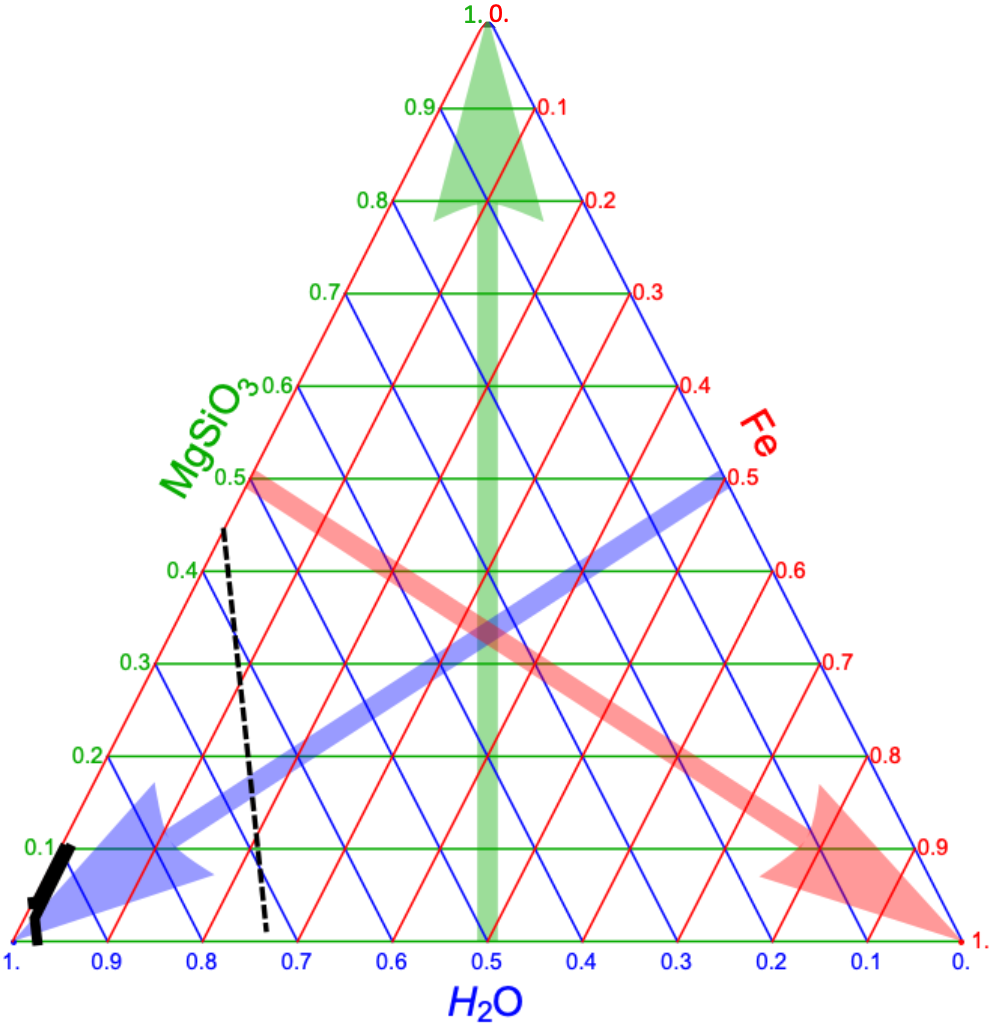}}
\end{subfigure}
\begin{subfigure}[\hdb]{
\includegraphics[width=0.45\textwidth]{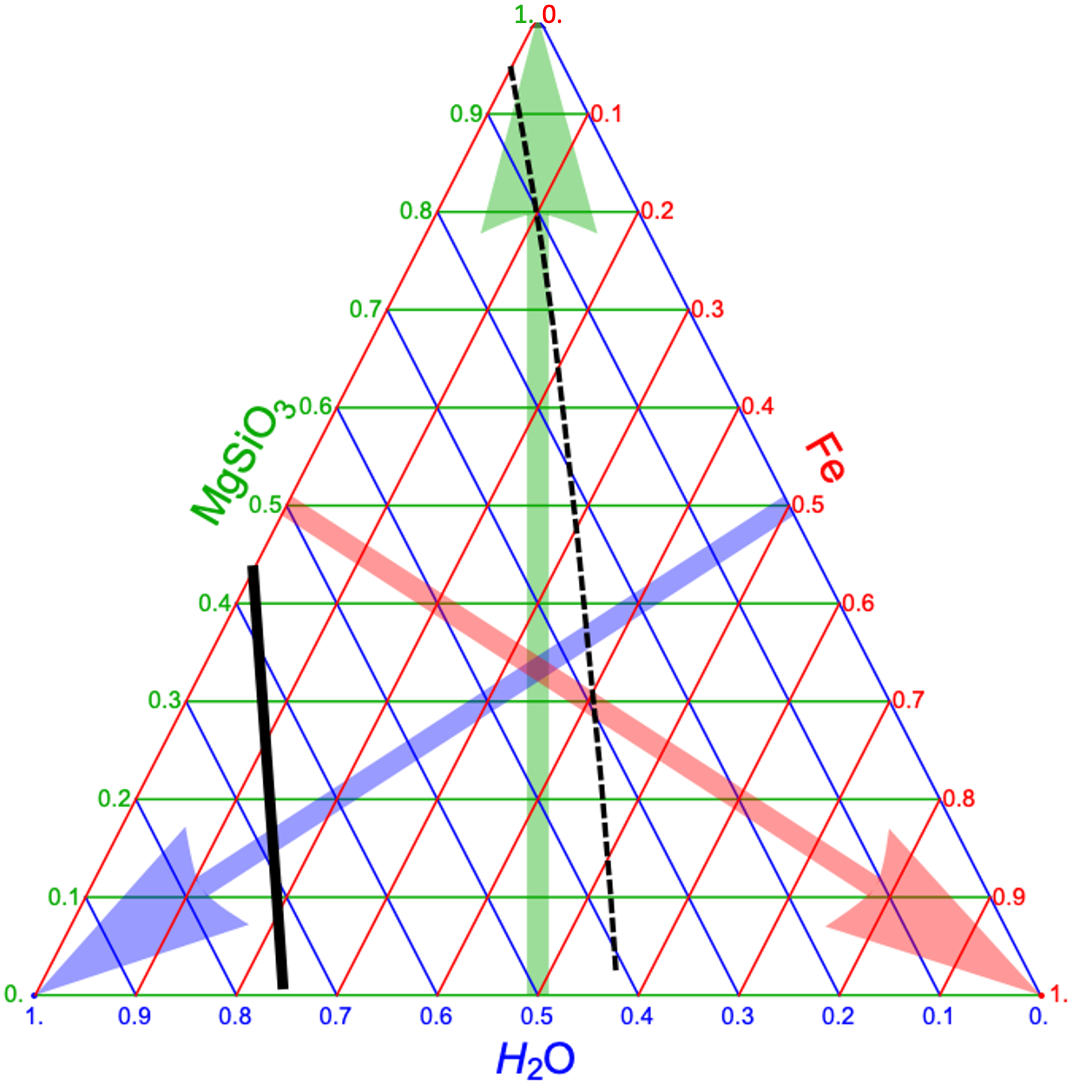}}
\end{subfigure}
\caption{\label{fig:water} Ternary diagrams using a 3-component H$_2$O/MgSiO$_3$/Fe model \citep{Zeng2013,Zeng2016}. The solid line outlines the median mass and radius of each planet, while the dashed line(s) delineate the 1$\sigma$ contours. At any point in the diagram the mass fractions can be found by following the three thin colored lines towards their respective side. }
\end{figure*}

To further investigate the potential $f_{H_2O}$, we explore three component models of H$_2$O, MgSiO$_3$, and Fe for four of our planets, excluding \hdc\ as its low density is inconsistent with these models. 
We use a numerical tool\footnote{\url{https://www.cfa.harvard.edu/~lzeng}} in order to solve for the interior structure of each planet and produce ternary diagrams of the range of combinations of MgSiO$_3$, Fe, and H$_2$O mass fractions that are compatible with the observed mass and radius \citep{Zeng2013,Zeng2016}. These ternary diagrams are shown in Figure~\ref{fig:water}. \gjb\ and \gjc\ both have a low H$_2$O fraction ($\leq$40\%) and a wide range of possibilities for MgSiO$_3$ \& Fe. \gjd\ is consistent with a high H$_2$O fraction (50--100\%) and small fractions of MgSiO$_3$ (0--50\%) and Fe (0--30\%). \hdb\ is consistent with a wide range for all three components (10--100\% H$_2$O, 0--90\% MgSiO$_3$, and 0--60\% Fe).

Both \citet{PrietoArranz2018} and \citet{Rice2019} suggest that photoevaporation may have sculpted the inner two rocky \gj\ planets. However, the outer planet, \gjd, must have retained a moderate fraction of volatiles to be consistent with its mass and radius. We examine whether the system as a whole is consistent with the theory of photoevaporation through calculating the minimum mass required of planet d to retain its atmosphere assuming planets b and c lost theirs to photoevaporation, as described in \citet{Owen2019}. We find the minimum mass for \gjd\ is 1 \mearth, lower than its mass of 3.3 \mearth. Therefore, this system is in agreement with this photevaporation model \citep{Owen2013,Owen2017}. Although, \gjd\ may have had a different type of atmospheric evolution other than photoevaporation. \citet{Kasper2020}
set stringent limits on the presence of any extended atmosphere around \gjd\ via high-resolution spectroscopy of the metastable 10,833 \mbox{\normalfont\AA} He triplet, inconsistent with current models of atmospheric formation and mass loss.

Furthermore, the three \gj\ planets span the radius gap at 1.7\rearth\ \citep{Fulton2017}. The inner two planets are high density and smaller than the radius gap (R$_b$=1.5\rearth, R$_c$=1.2\rearth) whereas the outer planet is lower density and larger than the radius gap (R$_d$=2.0\rearth). \hdb\ and c are both lower density and larger than the radius gap (R$_b$=2.4\rearth, R$_c$=4.4\rearth).
The five planets in these systems agree with a theory that planets smaller than 1.7\rearth\ are primarily composed of rocky cores and larger planets have additional volatile material that contributes to their radii \citep{Weiss2016,Fulton2017}.

\section{Conclusion}
\label{sec:conclusion}
In this paper, we characterized two systems, \hd\ and \gj. These bright stars host super-Earth and sub-Neptune planets well suited for atmospheric characterization by HST and JWST. From our Spitzer analysis (Section~\ref{sec:spitzer}) we improved the planets' ephemerides, enabling accurate transit prediction required for future atmospheric characterization through transmission spectroscopy. We incorporated Gaia parallaxes to update the stellar parameters for both systems and further constrained the limiting magnitude of nearby companions to \hd\ through imaging data (Section~\ref{sec:stellarchar}).

As the results of a multi-year high-cadence observing campaign with Keck/HIRES and Magellan/PFS, we improved the planets’ mass measurements in preparation for the interpretation of HST transmission spectra.
We measured planet masses in the \gj\ system to be M$_b$=$4.87\pm 0.37$ \mearth, M$_c$=$1.92\pm 0.49$ \mearth, and M$_d$=$3.42\pm 0.62$ \mearth. 
For \hd, we found planet masses of M$_b$=$10.5\pm 3.1$ \mearth\ and M$_c$=$12.0\pm3.8$ \mearth.
Atmospheric characterization of small planets benefits from mass detections at 5$\sigma$ significance \citep{Batalha2019}. We have achieved 5$\sigma$ masses for two planets with pending HST analyses, \gjb\ and \gjd\ (Hedges et al. in prep, Benneke et al in prep), and a 4$\sigma$ mass for the third, \hdc\ \citep{Kreidberg2020}. 

For \gj, stellar activity signatures in the photometry and Calcium II H\&K lines (Section~\ref{sec:stellaractivity}) informed our use of a Gaussian process to account for this activity in our radial velocity fit. We did not adopt the Gaussian process fit for our \hd\ analysis due to the higher AIC value and the lack of activity signatures seen in the Calcium II H\&K lines and radial velocity data. Hotter stars (T$_{\rm eff}>$6200 K) often have shallow convective envelopes and inefficient magnetic dynamos which result in fewer spots on the stellar surface \citep{Kraft1967}. Therefore, hotter stars like \hd\ may not have enough starspots for this type of Gaussian process to be effective.

We additionally explored the possible eccentricities for these planets through stability arguments. We found that low eccentricities are required for stability for these two closely-packed systems. 
We finally compared our derived masses and densities with previously published models to investigate interior compositions for these planets. 
We found \gjb\ and \gjc\ are both consistent with a 50/50 rock-iron composition, \gjd\ and \hdb\ both require additional volatiles, and \hdc\ is consistent with a $\sim$10\% by mass hydrogen/helium envelope.

\section{Appendix}
\startlongtable
\begin{deluxetable*}{lrrccc}
\tablecaption{\gj\ Radial Velocities \label{tab:gjrvs}}
\tablehead{
 \colhead{Time} & \colhead{RV} & \colhead{RV Unc.} & \colhead{S$_{\rm HK}$} & \colhead{H-alpha} & \colhead{Instrument} \\ \colhead{($BJD_{TDB}$)} & \colhead{(m s$^{-1}$)} & \colhead{(m s$^{-1}$)} & \colhead{} & \colhead{} & \colhead{}
} 
\startdata
2458787.89755 & 4.67 & 1.16 & 0.5741 & 0.05581 & HIRES \\
2458118.80405 & -0.01 & 1.08 & 0.7945 & 0.05629 & HIRES \\
2458646.10457 & 0.93 & 0.98 & 0.6726 & 0.05624 & HIRES \\
2458776.89384 & -4.37 & 1.61 & 0.5287 & 0.05622 & HIRES \\
2458125.76818 & -9.47 & 1.20 & 0.7162 & 0.05662 & HIRES \\
2458324.04450 & -1.40 & 1.22 & 0.6486 & 0.05680 & HIRES \\
2458300.01918 & -10.13 & 1.26 & 0.7221 & 0.05711 & HIRES \\
2458391.98260 & -13.69 & 1.21 & 0.6795 & 0.05708 & HIRES \\
2458476.72361 & -9.02 & 1.07 & 0.7117 & 0.05646 & HIRES \\
2458341.05653 & -4.39 & 1.14 & 0.6928 & 0.05671 & HIRES \\
2458264.10151 & -4.81 & 1.04 & 0.6822 & 0.05618 & HIRES \\
2458663.09765 & 1.91 & 1.04 & 0.6394 & 0.05695 & HIRES \\
2458361.06735 & -0.48 & 1.15 & 0.6929 & 0.05552 & HIRES \\
2458018.89464 & 4.00 & 1.14 & 0.6342 & 0.05697 & HIRES \\
2458462.76041 & 3.12 & 1.07 & 0.7086 & 0.05636 & HIRES \\
2458345.11938 & 0.04 & 1.46 & 0.6418 & 0.05576 & HIRES \\
2458285.11926 & 0.87 & 1.18 & 0.6411 & 0.05655 & HIRES \\
2458019.90188 & -4.03 & 1.13 & 0.6190 & 0.05611 & HIRES \\
2458724.91297 & -0.24 & 1.16 & 0.5269 & 0.05645 & HIRES \\
2458716.08123 & -4.38 & 1.30 & 0.5279 & 0.05626 & HIRES \\
2458396.85939 & -0.81 & 1.11 & 0.6396 & 0.05751 & HIRES \\
2458389.01961 & -0.55 & 1.40 & 0.6611 & 0.05627 & HIRES \\
2458346.10048 & -2.88 & 1.21 & 0.6692 & 0.05640 & HIRES \\
2458662.08768 & 0.41 & 1.02 & 0.6354 & 0.05657 & HIRES \\
2458724.02187 & -1.99 & 1.07 & 0.5176 & 0.05619 & HIRES \\
2458395.95532 & 3.94 & 1.01 & 0.6920 & 0.05704 & HIRES \\
2458124.79535 & -7.35 & 1.53 & 0.7996 & 0.05669 & HIRES \\
2458746.98946 & 0.61 & 1.06 & 0.5388 & 0.05542 & HIRES \\
2458709.89709 & -14.83 & 1.23 & 0.5539 & 0.05526 & HIRES \\
2458443.86005 & -7.99 & 1.11 & 0.7280 & 0.05749 & HIRES \\
2458295.07494 & 0.69 & 1.07 & 0.6887 & 0.05646 & HIRES \\
2458265.11117 & -7.06 & 1.29 & 0.6877 & 0.05642 & HIRES \\
2458733.03468 & -3.91 & 1.09 & 0.5733 & 0.05525 & HIRES \\
2458723.08397 & -9.46 & 1.26 & 0.5221 & 0.05615 & HIRES \\
2458436.77435 & -6.51 & 1.11 & -- & 0.05671 & HIRES \\
2458299.10180 & -4.48 & 1.19 & 0.7032 & 0.05699 & HIRES \\
2458091.81565 & 4.65 & 1.17 & 0.8633 & 0.05655 & HIRES \\
2458833.76746 & -2.47 & 1.16 & 0.5241 & 0.05508 & HIRES \\
2458393.94943 & -2.99 & 1.19 & 0.6665 & 0.05718 & HIRES \\
2458383.99765 & 4.71 & 1.24 & 0.6485 & 0.05708 & HIRES \\
2458309.01867 & -4.57 & 1.12 & 0.7225 & 0.05698 & HIRES \\
2458680.01498 & -6.33 & 1.08 & 0.6814 & 0.05561 & HIRES \\
2458832.81639 & 0.24 & 1.12 & 0.5802 & 0.05560 & HIRES \\
2458652.11022 & -3.33 & 1.07 & 0.6355 & 0.05712 & HIRES \\
2458370.03794 & -1.95 & 1.14 & 0.7151 & 0.05705 & HIRES \\
2458490.70951 & -1.38 & 1.25 & 0.7028 & 0.05679 & HIRES \\
2458350.05809 & 3.53 & 1.31 & 0.6635 & 0.05667 & HIRES \\
2458116.71383 & 4.52 & 1.08 & 0.6439 & 0.05704 & HIRES \\
2458293.10940 & -1.28 & 1.13 & 0.6735 & 0.05647 & HIRES \\
2458651.10778 & 0.07 & 1.03 & 0.6294 & 0.05590 & HIRES \\
2458802.84282 & -6.52 & 1.25 & 0.5617 & 0.05507 & HIRES \\
2458737.90776 & -5.52 & 1.11 & 0.5445 & 0.05570 & HIRES \\
2458364.04552 & -7.36 & 1.14 & 0.7050 & 0.05531 & HIRES \\
2458387.97876 & 1.60 & 1.08 & 0.6494 & 0.05737 & HIRES \\
2458844.80960 & 1.47 & 1.06 & 0.5513 & 0.05582 & HIRES \\
2458856.74435 & -6.18 & 1.46 & 0.5642 & 0.05665 & HIRES \\
2458827.80197 & -1.10 & 0.99 & 0.5579 & 0.03639 & HIRES \\
2458337.10349 & -0.30 & 1.15 & 0.7241 & 0.05619 & HIRES \\
2458715.07051 & 0.56 & 1.11 & 0.5301 & 0.05594 & HIRES \\
2458024.02110 & -1.63 & 1.34 & 0.6228 & 0.05587 & HIRES \\
2458491.71209 & -0.28 & 1.02 & 0.7160 & 0.05646 & HIRES \\
2458117.80128 & 4.23 & 1.07 & 0.7354 & 0.05637 & HIRES \\
2458306.03964 & -3.06 & 1.15 & 0.7271 & 0.05795 & HIRES \\
2458739.06355 & -4.30 & 1.18 & 0.5273 & 0.05488 & HIRES \\
2458099.72693 & -3.67 & 1.17 & 0.6767 & 0.05645 & HIRES \\
2458855.74402 & -10.15 & 1.39 & 0.5894 & 0.05608 & HIRES \\
2458845.75181 & 1.34 & 1.11 & 0.5558 & 0.05581 & HIRES \\
2458647.11328 & -9.14 & 0.93 & 0.6622 & 0.05659 & HIRES \\
2458819.83041 & 0.87 & 1.26 & 0.5522 & 0.05540 & HIRES \\
2458392.97702 & -7.79 & 1.11 & 0.6593 & 0.05694 & HIRES \\
2458328.93752 & -4.21 & 1.31 & 0.6909 & 0.05687 & HIRES \\
2458351.07086 & 1.74 & 1.35 & 0.6618 & 0.05672 & HIRES \\
2458329.99590 & -1.30 & 1.20 & 0.7188 & 0.05677 & HIRES \\
2458815.82726 & 0.04 & 1.13 & 0.5760 & 0.05550 & HIRES \\
2458324.95128 & 4.55 & 1.25 & 0.6582 & 0.05667 & HIRES \\
2458720.07172 & 2.33 & 1.33 & 0.5106 & 0.05586 & HIRES \\
2458367.01976 & 3.63 & 1.36 & 0.7497 & 0.05757 & HIRES \\
2458149.72124 & 7.73 & 1.23 & 0.8321 & 0.05668 & HIRES \\
2458797.92260 & -6.80 & 1.17 & 0.5685 & 0.05540 & HIRES \\
2458097.78803 & 2.71 & 1.13 & 0.6741 & 0.05631 & HIRES \\
2458291.10956 & -0.65 & 0.96 & 0.6571 & 0.05603 & HIRES \\
2458296.04279 & 7.36 & 0.97 & 0.7101 & 0.05693 & HIRES \\
2458338.10518 & 2.52 & 1.25 & 0.7101 & 0.05613 & HIRES \\
2458098.81794 & 7.06 & 1.17 & 0.6766 & 0.05621 & HIRES \\
2458389.98455 & -0.19 & 1.21 & 0.6644 & 0.05592 & HIRES \\
2458267.11611 & 6.21 & 1.11 & 0.6743 & 0.05696 & HIRES \\
2458292.10445 & -0.10 & 0.99 & 0.6650 & 0.05657 & HIRES \\
2458301.00170 & -9.07 & 1.26 & 0.7224 & 0.05679 & HIRES \\
2458366.10026 & 1.07 & 1.27 & 0.7132 & 0.05513 & HIRES \\
2458327.93655 & -5.02 & 1.23 & 0.6860 & 0.05686 & HIRES \\
2458266.11293 & 6.35 & 1.13 & 0.6764 & 0.05652 & HIRES \\
2458020.90394 & -7.94 & 1.13 & 0.6277 & 0.05662 & HIRES
\enddata
\tablenotetext{}{HIRES S$_{\rm HK}$ values have an uncertainty of 0.001.}
\end{deluxetable*}

\startlongtable
\begin{deluxetable*}{lrrccc}
\tablecaption{\hd\ Radial Velocities \label{tab:hdrvs}}
\tablehead{
 \colhead{Time} & \colhead{RV} & \colhead{RV Unc.} & \colhead{S$_{\rm HK}$} & \colhead{H-alpha} & \colhead{Instrument} \\ \colhead{($BJD_{TDB}$)} & \colhead{(m s$^{-1}$)} & \colhead{(m s$^{-1}$)} & \colhead{} & \colhead{} & \colhead{}
}
\startdata
2457746.13882 & -6.58 & 4.11 & 0.1392 & 0.03299 & HIRES \\
2457746.14353 & -3.36 & 4.01 & 0.1391 & 0.03294 & HIRES \\
2457747.06934 & 0.11 & 3.79 & 0.1398 & 0.03288 & HIRES \\
2457747.10551 & 1.64 & 4.17 & 0.1399 & 0.03321 & HIRES \\
2457747.15981 & 15.16 & 3.95 & 0.1404 & 0.03313 & HIRES \\
2457760.09582 & 2.62 & 4.15 & 0.1369 & 0.03307 & HIRES \\
2457760.13104 & -13.59 & 3.97 & 0.1375 & 0.03306 & HIRES \\
2457760.17348 & -7.50 & 4.13 & 0.1398 & 0.03304 & HIRES \\
2457764.01751 & 6.98 & 4.29 & 0.1374 & 0.03304 & HIRES \\
2457764.05279 & 1.81 & 4.50 & 0.1386 & 0.03283 & HIRES \\
2457764.09032 & 2.78 & 4.01 & 0.1395 & 0.03307 & HIRES \\
2457764.09369 & 3.45 & 3.94 & 0.1392 & 0.03305 & HIRES \\
2457764.09704 & 8.54 & 3.92 & 0.1390 & 0.03305 & HIRES \\
2457764.13272 & -10.27 & 4.58 & 0.1396 & 0.03316 & HIRES \\
2457764.17257 & 5.27 & 3.83 & 0.1386 & 0.03320 & HIRES \\
2457765.02368 & -10.26 & 3.91 & 0.1387 & 0.03288 & HIRES \\
2457765.02889 & -6.80 & 4.20 & 0.1382 & 0.03293 & HIRES \\
2457765.03277 & -5.33 & 4.16 & 0.1389 & 0.03288 & HIRES \\
2457765.06829 & -7.55 & 3.94 & 0.1382 & 0.03296 & HIRES \\
2457765.14462 & -1.96 & 3.95 & 0.1379 & 0.03316 & HIRES \\
2457765.15150 & -7.33 & 4.13 & 0.1386 & 0.03325 & HIRES \\
2457765.15892 & -0.70 & 4.17 & 0.1380 & 0.03313 & HIRES \\
2457766.02041 & 0.03 & 4.10 & 0.1356 & 0.03301 & HIRES \\
2457766.05479 & -15.51 & 4.32 & 0.1362 & 0.03296 & HIRES \\
2457766.10347 & -17.81 & 4.16 & 0.1371 & 0.03334 & HIRES \\
2457766.13313 & -12.24 & 4.04 & 0.1371 & 0.03325 & HIRES \\
2457766.17504 & -16.04 & 4.19 & 0.1364 & 0.03330 & HIRES \\
2457775.00337 & -18.93 & 5.04 & 0.1395 & 0.03277 & HIRES \\
2457775.08336 & -7.43 & 5.14 & 0.1382 & 0.03290 & HIRES \\
2457775.14543 & 15.95 & 5.16 & 0.1386 & 0.03290 & HIRES \\
2457775.17945 & 13.34 & 5.29 & 0.1318 & 0.03282 & HIRES \\
2457775.97301 & 6.76 & 5.29 & 0.1366 & 0.03331 & HIRES \\
2457776.03370 & -0.88 & 5.29 & 0.1371 & 0.03312 & HIRES \\
2457776.07307 & -3.10 & 5.30 & 0.1377 & 0.03291 & HIRES \\
2457776.11667 & -0.40 & 5.40 & 0.1358 & 0.03291 & HIRES \\
2457776.17591 & 9.27 & 4.93 & 0.1350 & 0.03307 & HIRES \\
2457788.03576 & -6.97 & 5.33 & 0.1337 & 0.03292 & HIRES \\
2457788.09236 & -8.83 & 5.18 & 0.1353 & 0.03288 & HIRES \\
2457788.14459 & 10.39 & 5.43 & 0.1361 & 0.03276 & HIRES \\
2457788.96764 & -1.52 & 4.97 & 0.1368 & 0.03274 & HIRES \\
2457789.03425 & -11.90 & 5.20 & 0.1358 & 0.03308 & HIRES \\
2457789.07579 & -16.63 & 4.92 & 0.1356 & 0.03296 & HIRES \\
2457789.12552 & -1.69 & 4.93 & 0.1309 & 0.03305 & HIRES \\
2457789.93588 & -14.19 & 5.35 & 0.1372 & 0.03276 & HIRES \\
2457789.97055 & -6.69 & 5.46 & 0.1371 & 0.03302 & HIRES \\
2457790.02625 & 10.37 & 4.80 & 0.1380 & 0.03304 & HIRES \\
2457790.07667 & 11.05 & 5.24 & 0.1374 & 0.03305 & HIRES \\
2457790.11637 & -0.66 & 5.88 & 0.1375 & 0.03307 & HIRES \\
2457790.94126 & 3.04 & 4.48 & 0.1340 & 0.03293 & HIRES \\
2457790.98855 & 3.61 & 4.85 & 0.1321 & 0.03305 & HIRES \\
2457791.02903 & 3.53 & 4.62 & 0.1304 & 0.03301 & HIRES \\
2457791.06239 & 3.40 & 4.51 & 0.1316 & 0.03311 & HIRES \\
2457791.13144 & -5.63 & 4.45 & 0.1331 & 0.03314 & HIRES \\
2457792.95306 & 5.29 & 4.31 & 0.1331 & 0.03268 & HIRES \\
2457793.01216 & 1.33 & 4.48 & 0.1349 & 0.03287 & HIRES \\
2457793.06477 & -1.47 & 4.64 & 0.1353 & 0.03306 & HIRES \\
2457793.09752 & 7.74 & 5.09 & 0.1331 & 0.03315 & HIRES \\
2457794.01892 & -1.13 & 4.53 & 0.1361 & 0.03299 & HIRES \\
2457794.06873 & 9.90 & 4.53 & 0.1362 & 0.03300 & HIRES \\
2457794.12856 & -9.77 & 4.72 & 0.1354 & 0.03300 & HIRES \\
2457794.96285 & 2.04 & 4.70 & 0.1356 & 0.03299 & HIRES \\
2457795.00019 & 14.61 & 4.38 & 0.1358 & 0.03312 & HIRES \\
2457795.11828 & 3.57 & 4.55 & 0.1356 & 0.03295 & HIRES \\
2457796.00198 & 17.72 & 5.51 & 0.1333 & 0.03305 & HIRES \\
2457802.91637 & -1.14 & 4.26 & 0.1358 & 0.03303 & HIRES \\
2457802.94277 & 0.19 & 4.26 & 0.1354 & 0.03316 & HIRES \\
2457803.89893 & -10.62 & 4.28 & 0.1332 & 0.03302 & HIRES \\
2457803.92126 & -5.97 & 4.48 & 0.1356 & 0.03288 & HIRES \\
2457804.89453 & 1.83 & 8.27 & 0.1174 & 0.03349 & HIRES \\
2457805.87966 & -16.14 & 4.06 & 0.1360 & 0.03259 & HIRES \\
2457805.91546 & 3.79 & 4.77 & 0.1385 & 0.03309 & HIRES \\
2457805.94644 & -16.41 & 5.74 & 0.1370 & 0.03288 & HIRES \\
2457806.87926 & 0.21 & 5.35 & 0.1304 & 0.03285 & HIRES \\
2457806.91553 & -7.60 & 4.51 & 0.1348 & 0.03298 & HIRES \\
2457806.96445 & -8.99 & 4.41 & 0.1355 & 0.03353 & HIRES \\
2457828.91583 & -15.48 & 4.56 & 0.1359 & 0.03302 & HIRES \\
2457828.97000 & -1.71 & 4.68 & 0.1355 & 0.03307 & HIRES \\
2457829.04172 & -10.00 & 4.83 & 0.1359 & 0.03298 & HIRES \\
2457829.83091 & -9.99 & 4.62 & 0.1350 & 0.03283 & HIRES \\
2457829.94218 & -16.57 & 4.63 & 0.1343 & 0.03296 & HIRES \\
2457830.05077 & -9.15 & 4.37 & 0.1360 & 0.03296 & HIRES \\
2457830.95834 & 3.17 & 4.85 & 0.1376 & 0.03369 & HIRES \\
2457830.97649 & -11.76 & 4.62 & 0.1373 & 0.03382 & HIRES \\
2457831.02642 & -13.07 & 4.52 & 0.1364 & 0.03318 & HIRES \\
2457886.92496 & -11.37 & 3.97 & 0.1365 & 0.03352 & HIRES \\
2457887.94831 & -11.08 & 4.43 & 0.1361 & 0.03325 & HIRES \\
2457887.97540 & -4.90 & 4.18 & 0.1324 & 0.03308 & HIRES \\
2457925.75207 & 2.27 & 4.26 & 0.1332 & 0.03285 & HIRES \\
2457925.75570 & -0.51 & 3.95 & 0.1340 & 0.03281 & HIRES \\
2457925.75951 & -5.71 & 4.27 & 0.1336 & 0.03282 & HIRES \\
2457925.84733 & -5.79 & 3.96 & 0.1355 & 0.03266 & HIRES \\
2457925.85224 & 6.53 & 3.72 & 0.1351 & 0.03262 & HIRES \\
2457925.85715 & -11.74 & 4.20 & 0.1351 & 0.03268 & HIRES \\
2457925.87848 & 2.19 & 4.05 & 0.1353 & 0.03260 & HIRES \\
2457926.75519 & 5.55 & 4.06 & 0.1362 & 0.03313 & HIRES \\
2457926.75814 & 0.27 & 4.01 & 0.1356 & 0.03312 & HIRES \\
2457926.76117 & 4.34 & 4.36 & 0.1359 & 0.03312 & HIRES \\
2457926.81714 & -3.79 & 4.44 & 0.1351 & 0.03279 & HIRES \\
2457926.82071 & -6.63 & 4.53 & 0.1351 & 0.03279 & HIRES \\
2457926.82430 & -6.60 & 4.53 & 0.1351 & 0.03284 & HIRES \\
2457926.87381 & 1.71 & 4.09 & 0.1351 & 0.03257 & HIRES \\
2457926.87805 & -13.30 & 4.41 & 0.1349 & 0.03260 & HIRES \\
2457926.88191 & -12.34 & 4.85 & 0.1349 & 0.03266 & HIRES \\
2457932.74875 & -15.49 & 4.25 & 0.1338 & 0.03287 & HIRES \\
2457932.75265 & -7.25 & 4.22 & 0.1334 & 0.03286 & HIRES \\
2457932.75644 & -7.54 & 4.53 & 0.1337 & 0.03304 & HIRES \\
2457932.82664 & -7.59 & 4.14 & 0.1327 & 0.03280 & HIRES \\
2457932.83122 & -20.21 & 4.66 & 0.1328 & 0.03282 & HIRES \\
2457932.83577 & 2.63 & 4.09 & 0.1329 & 0.03278 & HIRES \\
2457939.75845 & 2.40 & 4.46 & 0.1273 & 0.03304 & HIRES \\
2457939.76412 & 4.70 & 4.70 & 0.1259 & 0.03309 & HIRES \\
2457939.76959 & 3.76 & 4.65 & 0.1278 & 0.03301 & HIRES \\
2457940.79054 & 6.71 & 4.28 & 0.1333 & 0.03268 & HIRES \\
2457940.79493 & -4.74 & 4.22 & 0.1321 & 0.03273 & HIRES \\
2457940.79981 & -6.94 & 4.08 & 0.1302 & 0.03280 & HIRES \\
2457964.75862 & -2.12 & 4.56 & 0.1350 & 0.03332 & HIRES \\
2457964.76469 & -1.59 & 4.59 & 0.1351 & 0.03317 & HIRES \\
2458113.08369 & -8.93 & 4.71 & 0.1360 & 0.03293 & HIRES \\
2458113.09308 & -8.40 & 4.12 & 0.1366 & 0.03296 & HIRES \\
2458113.09715 & -5.69 & 4.00 & 0.1362 & 0.03305 & HIRES \\
2458114.03130 & 0.92 & 4.34 & 0.1372 & 0.03288 & HIRES \\
2458114.03639 & -11.77 & 4.31 & 0.1317 & 0.03292 & HIRES \\
2458114.04179 & -6.28 & 4.43 & 0.1341 & 0.03297 & HIRES \\
2458114.08344 & -7.93 & 4.44 & 0.1381 & 0.03321 & HIRES \\
2458114.08768 & -8.53 & 4.13 & 0.1372 & 0.03322 & HIRES \\
2458114.09184 & 3.60 & 4.22 & 0.1381 & 0.03335 & HIRES \\
2458149.95743 & -7.48 & 4.10 & 0.1389 & 0.03304 & HIRES \\
2458149.96341 & -3.46 & 4.12 & 0.1379 & 0.03307 & HIRES \\
2458149.96889 & -4.02 & 4.30 & 0.1384 & 0.03307 & HIRES \\
2458150.10402 & -5.04 & 4.86 & 0.1271 & 0.03348 & HIRES \\
2458150.11108 & -2.19 & 4.73 & 0.1338 & 0.03365 & HIRES \\
2458150.11674 & -0.77 & 4.72 & 0.1351 & 0.03353 & HIRES \\
2458150.93670 & -5.13 & 4.53 & 0.1371 & 0.03294 & HIRES \\
2458150.94605 & -1.17 & 4.89 & 0.1380 & 0.03301 & HIRES \\
2458150.95438 & -8.99 & 4.76 & 0.1376 & 0.03298 & HIRES \\
2458151.01303 & 5.26 & 4.34 & 0.1378 & 0.03314 & HIRES \\
2458151.01794 & 4.91 & 4.90 & 0.1379 & 0.03306 & HIRES \\
2458151.02280 & 3.30 & 4.11 & 0.1378 & 0.03315 & HIRES \\
2458161.11153 & 1.27 & 4.29 & 0.1358 & 0.03284 & HIRES \\
2458161.11535 & -3.61 & 4.58 & 0.1364 & 0.03298 & HIRES \\
2458161.11932 & -14.38 & 4.54 & 0.1355 & 0.03291 & HIRES \\
2458194.96586 & 3.35 & 5.03 & 0.1396 & 0.03343 & HIRES \\
2458194.96962 & 0.96 & 4.92 & 0.1395 & 0.03331 & HIRES \\
2458194.97338 & 5.12 & 4.76 & 0.1397 & 0.03339 & HIRES \\
2458199.95986 & -11.36 & 4.80 & 0.1371 & 0.04975 & HIRES \\
2458247.95188 & -0.95 & 4.39 & 0.1344 & 0.03361 & HIRES \\
2458247.98535 & -6.99 & 4.91 & 0.1342 & 0.03202 & HIRES \\
2458284.74670 & -7.79 & 4.25 & 0.1378 & 0.03300 & HIRES \\
2458284.75052 & -7.46 & 4.53 & 0.1379 & 0.03303 & HIRES \\
2458284.75435 & -0.95 & 4.44 & 0.1388 & 0.03302 & HIRES \\
2458294.75288 & -12.54 & 4.11 & 0.1359 & 0.03314 & HIRES \\
2458294.75633 & -7.78 & 4.27 & 0.1359 & 0.03319 & HIRES \\
2458294.75966 & -5.97 & 4.18 & 0.1361 & 0.03320 & HIRES \\
2458295.76177 & -1.04 & 4.64 & 0.1347 & 0.03331 & HIRES \\
2458295.76762 & -4.38 & 3.85 & 0.1354 & 0.03329 & HIRES \\
2458295.77317 & -12.32 & 4.34 & 0.1362 & 0.03334 & HIRES \\
2458298.76417 & 5.58 & 4.60 & 0.1363 & 0.03332 & HIRES \\
2458298.76812 & -4.93 & 4.35 & 0.1366 & 0.03328 & HIRES \\
2458298.77221 & 0.53 & 4.39 & 0.1365 & 0.03322 & HIRES \\
2458299.75015 & -10.27 & 4.37 & 0.1376 & 0.03318 & HIRES \\
2458299.75451 & -2.26 & 4.75 & 0.1378 & 0.03316 & HIRES \\
2458299.75887 & -2.35 & 4.22 & 0.1379 & 0.03314 & HIRES \\
2458300.76210 & -7.52 & 4.50 & 0.1355 & 0.03318 & HIRES \\
2458300.76556 & -2.26 & 4.31 & 0.1360 & 0.03310 & HIRES \\
2458300.76895 & 1.77 & 4.38 & 0.1355 & 0.03325 & HIRES \\
2458301.77150 & 17.34 & 4.99 & 0.1389 & 0.03365 & HIRES \\
2458303.74928 & -4.68 & 4.28 & 0.1381 & 0.03305 & HIRES \\
2458303.75263 & 1.69 & 4.24 & 0.1380 & 0.03304 & HIRES \\
2458303.75621 & 3.79 & 4.03 & 0.1381 & 0.03306 & HIRES \\
2458305.81046 & -4.01 & 4.26 & 0.1355 & 0.03326 & HIRES \\
2458305.81521 & -9.77 & 4.32 & 0.1350 & 0.03319 & HIRES \\
2458305.82007 & -9.03 & 4.35 & 0.1346 & 0.03314 & HIRES \\
2458307.77651 & -13.55 & 4.72 & 0.1319 & 0.03310 & HIRES \\
2458307.78206 & -8.90 & 4.51 & 0.1336 & 0.03316 & HIRES \\
2458307.78721 & -16.93 & 4.61 & 0.1328 & 0.03301 & HIRES \\
2458308.80024 & -11.53 & 4.26 & 0.1346 & 0.03299 & HIRES \\
2458308.80495 & -15.77 & 4.50 & 0.1344 & 0.03271 & HIRES \\
2458308.80947 & -12.66 & 4.17 & 0.1344 & 0.03296 & HIRES \\
2458323.75053 & -11.87 & 4.18 & 0.1322 & 0.03295 & HIRES \\
2458323.75826 & -20.17 & 4.47 & 0.1286 & 0.03299 & HIRES \\
2458323.76745 & -2.81 & 4.71 & 0.1270 & 0.03299 & HIRES \\
2458324.74711 & -15.96 & 4.37 & 0.1347 & 0.03298 & HIRES \\
2458324.75189 & -14.12 & 4.39 & 0.1339 & 0.03306 & HIRES \\
2458324.75797 & -16.04 & 4.22 & 0.1338 & 0.03293 & HIRES \\
2458491.06151 & 5.22 & 3.96 & 0.1398 & 0.03326 & HIRES \\
2458491.06721 & 6.10 & 3.89 & 0.1400 & 0.03319 & HIRES \\
2458491.07267 & -1.59 & 4.12 & 0.1394 & 0.03321 & HIRES \\
2458491.12711 & 0.38 & 3.96 & 0.1389 & 0.03323 & HIRES \\
2458491.13151 & -1.88 & 3.92 & 0.1382 & 0.03319 & HIRES \\
2458491.13662 & 0.09 & 3.95 & 0.1397 & 0.03314 & HIRES \\
2458492.00757 & 0.56 & 3.90 & 0.1389 & 0.03308 & HIRES \\
2458492.01111 & 8.38 & 4.13 & 0.1387 & 0.03305 & HIRES \\
2458492.01468 & -2.83 & 4.04 & 0.1392 & 0.03305 & HIRES \\
2458492.07044 & 19.16 & 4.18 & 0.1387 & 0.03330 & HIRES \\
2458492.07379 & -3.72 & 3.81 & 0.1390 & 0.03335 & HIRES \\
2458492.07719 & -2.58 & 3.66 & 0.1384 & 0.03334 & HIRES \\
2458492.12123 & -3.90 & 4.32 & 0.1393 & 0.03341 & HIRES \\
2458492.12472 & 8.06 & 4.02 & 0.1393 & 0.03335 & HIRES \\
2458492.12820 & 5.37 & 4.12 & 0.1391 & 0.03340 & HIRES \\
2458532.93369 & 9.71 & 4.16 & 0.1387 & 0.03312 & HIRES \\
2458532.93872 & -6.50 & 3.96 & 0.1382 & 0.03330 & HIRES \\
2458532.94371 & 1.36 & 4.06 & 0.1376 & 0.03325 & HIRES \\
2458533.00188 & 9.85 & 4.29 & 0.1381 & 0.03344 & HIRES \\
2458533.00611 & 7.75 & 4.01 & 0.1376 & 0.03338 & HIRES \\
2458533.01024 & 0.00 & 4.17 & 0.1373 & 0.03352 & HIRES \\
2458533.06987 & 6.77 & 4.05 & 0.1378 & 0.03354 & HIRES \\
2458533.07437 & 7.20 & 3.87 & 0.1379 & 0.03346 & HIRES \\
2458533.07857 & 5.43 & 4.30 & 0.1385 & 0.03354 & HIRES \\
2458559.86044 & -14.59 & 4.66 & 0.1407 & 0.03359 & HIRES \\
2458559.86545 & -17.11 & 4.51 & 0.1406 & 0.03338 & HIRES \\
2458559.86996 & -17.09 & 4.38 & 0.1401 & 0.03353 & HIRES \\
2458559.95743 & -5.17 & 4.54 & 0.1402 & 0.03396 & HIRES \\
2458559.96168 & -9.94 & 4.42 & 0.1409 & 0.03391 & HIRES \\
2458559.96622 & -2.16 & 4.41 & 0.1401 & 0.03379 & HIRES \\
2458560.01730 & -1.43 & 4.56 & 0.1404 & 0.03373 & HIRES \\
2458560.02263 & -3.45 & 4.24 & 0.1403 & 0.03365 & HIRES \\
2458560.02774 & 2.49 & 4.20 & 0.1402 & 0.03369 & HIRES \\
2458566.95939 & 2.30 & 4.68 & 0.1372 & 0.03338 & HIRES \\
2458566.96623 & 12.43 & 4.79 & 0.1372 & 0.03317 & HIRES \\
2458566.97435 & 27.78 & 4.88 & 0.1368 & 0.03342 & HIRES \\
2458567.02561 & 80.40 & 5.43 & 0.1383 & 0.03344 & HIRES \\
2458567.04056 & 102.65 & 5.85 & 0.1380 & 0.03333 & HIRES \\
2458567.04794 & 18.92 & 4.49 & 0.1367 & 0.03323 & HIRES \\
2458568.81903 & -6.03 & 4.33 & 0.1364 & 0.03310 & HIRES \\
2458568.82355 & 6.29 & 4.42 & 0.1361 & 0.03316 & HIRES \\
2458568.82793 & 1.12 & 4.82 & 0.1366 & 0.03312 & HIRES \\
2458568.91443 & 0.94 & 4.45 & 0.1358 & 0.03333 & HIRES \\
2458568.91785 & 3.22 & 4.70 & 0.1361 & 0.03342 & HIRES \\
2458568.92136 & -8.03 & 4.37 & 0.1355 & 0.03332 & HIRES \\
2458569.83303 & 12.47 & 4.63 & 0.1354 & 0.03323 & HIRES \\
2458569.83678 & 21.89 & 4.50 & 0.1356 & 0.03320 & HIRES \\
2458569.84071 & 0.70 & 4.49 & 0.1365 & 0.03308 & HIRES \\
2458569.92580 & 12.24 & 4.55 & 0.1351 & 0.03326 & HIRES \\
2458569.92944 & 12.05 & 4.39 & 0.1359 & 0.03329 & HIRES \\
2458569.93329 & 3.50 & 4.61 & 0.1357 & 0.03326 & HIRES \\
2458584.88918 & 4.05 & 4.72 & 0.1365 & 0.03332 & HIRES \\
2458584.89615 & -2.20 & 4.67 & 0.1366 & 0.03318 & HIRES \\
2458584.90306 & 4.27 & 4.81 & 0.1365 & 0.03321 & HIRES \\
2458592.95226 & -22.65 & 4.52 & 0.1364 & 0.03323 & HIRES \\
2458592.95676 & -23.53 & 4.49 & 0.1365 & 0.03324 & HIRES \\
2458592.96169 & -46.04 & 4.53 & 0.1359 & 0.03325 & HIRES \\
2458595.81757 & 4.50 & 4.60 & 0.1380 & 0.03305 & HIRES \\
2458595.82100 & -0.75 & 4.48 & 0.1376 & 0.03310 & HIRES \\
2458595.82449 & 5.94 & 4.84 & 0.1380 & 0.03307 & HIRES \\
2458595.87183 & 12.89 & 4.53 & 0.1376 & 0.03299 & HIRES \\
2458595.87533 & -17.78 & 4.92 & 0.1380 & 0.03295 & HIRES \\
2458595.87873 & 3.46 & 4.92 & 0.1377 & 0.03302 & HIRES \\
2458599.77326 & -2.12 & 4.51 & 0.1366 & 0.03295 & HIRES \\
2458599.77656 & -6.93 & 4.55 & 0.1369 & 0.03310 & HIRES \\
2458599.77994 & -1.36 & 4.67 & 0.1363 & 0.03310 & HIRES \\
2458610.86968 & -0.13 & 4.48 & 0.1379 & 0.03272 & HIRES \\
2458610.87354 & 7.16 & 4.34 & 0.1380 & 0.03264 & HIRES \\
2458610.87724 & 3.72 & 4.14 & 0.1382 & 0.03284 & HIRES \\
2458615.76217 & -0.99 & 4.49 & 0.1374 & 0.03339 & HIRES \\
2458615.76568 & 6.66 & 4.11 & 0.1381 & 0.03347 & HIRES \\
2458615.76915 & -0.59 & 4.24 & 0.1383 & 0.03333 & HIRES \\
2458615.84636 & 8.56 & 4.32 & 0.1378 & 0.03364 & HIRES \\
2458615.84979 & 1.18 & 4.49 & 0.1378 & 0.03375 & HIRES \\
2458615.85320 & -6.47 & 4.26 & 0.1378 & 0.03353 & HIRES \\
2458616.83882 & -5.61 & 3.96 & 0.1373 & 0.03290 & HIRES \\
2458616.84231 & -1.77 & 4.00 & 0.1376 & 0.03296 & HIRES \\
2458616.84587 & 5.76 & 4.50 & 0.1370 & 0.03301 & HIRES \\
2458616.89557 & -4.15 & 4.05 & 0.1378 & 0.03308 & HIRES \\
2458616.89872 & 6.75 & 4.13 & 0.1372 & 0.03305 & HIRES \\
2458616.90188 & 7.06 & 4.05 & 0.1369 & 0.03330 & HIRES \\
2458622.80874 & -4.91 & 4.37 & 0.1373 & 0.03326 & HIRES \\
2458622.81217 & -8.17 & 4.30 & 0.1376 & 0.03343 & HIRES \\
2458622.81565 & 6.23 & 4.19 & 0.1376 & 0.03344 & HIRES \\
2458622.88602 & 1.62 & 4.03 & 0.1364 & 0.03308 & HIRES \\
2458622.89232 & -5.79 & 4.14 & 0.1366 & 0.03342 & HIRES \\
2458622.89780 & -13.17 & 4.13 & 0.1368 & 0.03338 & HIRES \\
2458623.74265 & 10.84 & 4.03 & 0.1350 & 0.03318 & HIRES \\
2458623.74675 & -1.11 & 4.46 & 0.1357 & 0.03338 & HIRES \\
2458623.75081 & 11.45 & 4.33 & 0.1336 & 0.03342 & HIRES \\
2458623.86571 & -6.69 & 4.12 & 0.1369 & 0.03311 & HIRES \\
2458623.87014 & -4.93 & 4.09 & 0.1370 & 0.03307 & HIRES \\
2458623.87500 & -0.48 & 3.78 & 0.1370 & 0.03318 & HIRES \\
2458627.74481 & 0.33 & 4.12 & 0.1366 & 0.03356 & HIRES \\
2458627.74848 & -12.67 & 4.34 & 0.1369 & 0.03367 & HIRES \\
2458627.75218 & -10.21 & 4.58 & 0.1374 & 0.03350 & HIRES \\
2458627.84051 & -9.14 & 4.17 & 0.1382 & 0.03331 & HIRES \\
2458627.84511 & -6.16 & 4.22 & 0.1383 & 0.03334 & HIRES \\
2458627.84943 & -9.02 & 4.20 & 0.1384 & 0.03330 & HIRES \\
2458628.74062 & -9.19 & 4.16 & 0.1349 & 0.03301 & HIRES \\
2458628.74411 & 6.28 & 4.42 & 0.1362 & 0.03304 & HIRES \\
2458628.74779 & -7.19 & 4.28 & 0.1356 & 0.03308 & HIRES \\
2458628.81257 & -6.73 & 4.48 & 0.1374 & 0.03307 & HIRES \\
2458628.81667 & 4.30 & 4.14 & 0.1380 & 0.03327 & HIRES \\
2458628.82058 & 1.21 & 4.17 & 0.1377 & 0.03307 & HIRES \\
2458632.74785 & 0.10 & 4.11 & 0.1344 & 0.03318 & HIRES \\
2458632.75135 & 0.42 & 4.30 & 0.1342 & 0.03320 & HIRES \\
2458632.75491 & -1.10 & 4.16 & 0.1348 & 0.03320 & HIRES \\
2458632.85224 & -12.75 & 4.12 & 0.1339 & 0.03307 & HIRES \\
2458632.85554 & 3.56 & 4.24 & 0.1337 & 0.03296 & HIRES \\
2458632.85885 & -16.81 & 4.22 & 0.1345 & 0.03304 & HIRES \\
2458633.76878 & -1.49 & 4.40 & 0.1364 & 0.03313 & HIRES \\
2458633.77707 & 2.85 & 4.53 & 0.1363 & 0.03314 & HIRES \\
2458633.78548 & -5.73 & 4.16 & 0.1364 & 0.03303 & HIRES \\
2458633.82325 & -4.78 & 4.38 & 0.1363 & 0.03301 & HIRES \\
2458633.82716 & 3.52 & 4.45 & 0.1357 & 0.03300 & HIRES \\
2458633.83163 & -7.87 & 4.53 & 0.1360 & 0.03303 & HIRES \\
2458647.75057 & -6.88 & 4.57 & 0.1350 & 0.03340 & HIRES \\
2458647.75402 & -2.60 & 4.47 & 0.1360 & 0.03335 & HIRES \\
2458647.75759 & -4.60 & 3.77 & 0.1351 & 0.03331 & HIRES \\
2458647.82239 & 6.83 & 4.60 & 0.1361 & 0.03322 & HIRES \\
2458647.82634 & 0.37 & 4.18 & 0.1363 & 0.03314 & HIRES \\
2458647.83030 & -7.89 & 4.33 & 0.1362 & 0.03339 & HIRES \\
2458650.75680 & -9.52 & 4.33 & 0.1357 & 0.03304 & HIRES \\
2458650.76074 & -13.46 & 4.52 & 0.1359 & 0.03302 & HIRES \\
2458650.76468 & -0.39 & 4.52 & 0.1357 & 0.03304 & HIRES \\
2458650.84263 & -12.51 & 4.40 & 0.1367 & 0.03270 & HIRES \\
2458650.84601 & -11.60 & 4.64 & 0.1362 & 0.03267 & HIRES \\
2458650.84931 & -1.02 & 4.25 & 0.1370 & 0.03270 & HIRES \\
2458651.75457 & 0.14 & 4.62 & 0.1341 & 0.03348 & HIRES \\
2458651.75841 & 7.26 & 4.53 & 0.1347 & 0.03341 & HIRES \\
2458651.76254 & -11.04 & 4.43 & 0.1329 & 0.03342 & HIRES \\
2458651.81380 & -9.22 & 4.26 & 0.1361 & 0.03318 & HIRES \\
2458651.81814 & 6.26 & 4.34 & 0.1359 & 0.03317 & HIRES \\
2458651.82221 & -12.08 & 4.10 & 0.1357 & 0.03317 & HIRES \\
2458659.77465 & -15.10 & 4.48 & 0.1370 & 0.03322 & HIRES \\
2458659.77766 & -11.07 & 4.58 & 0.1368 & 0.03306 & HIRES \\
2458659.78086 & -13.61 & 4.21 & 0.1368 & 0.03306 & HIRES \\
2458660.76770 & 3.46 & 4.37 & 0.1361 & 0.03314 & HIRES \\
2458660.77131 & 12.67 & 4.38 & 0.1359 & 0.03303 & HIRES \\
2458660.77482 & 11.91 & 4.31 & 0.1371 & 0.03314 & HIRES \\
2458665.77561 & -5.12 & 4.17 & 0.1367 & 0.03339 & HIRES \\
2458665.77882 & -8.38 & 4.09 & 0.1370 & 0.03337 & HIRES \\
2458665.78206 & -6.86 & 4.18 & 0.1368 & 0.03345 & HIRES \\
2458679.77419 & 2.38 & 4.17 & 0.1377 & 0.03294 & HIRES \\
2458679.77729 & 3.83 & 4.12 & 0.1374 & 0.03288 & HIRES \\
2458679.78036 & 5.70 & 4.56 & 0.1373 & 0.03288 & HIRES \\
2458709.73744 & -8.95 & 5.03 & 0.1107 & 0.03287 & HIRES \\
2458809.13240 & -3.77 & 4.98 & 0.1236 & 0.03305 & HIRES \\
2458809.13605 & -4.17 & 4.80 & 0.1254 & 0.03298 & HIRES \\
2458809.13960 & -20.52 & 4.87 & 0.1250 & 0.03299 & HIRES \\
2458828.12545 & 9.33 & 4.22 & 0.1236 & 0.03318 & HIRES \\
2458828.12836 & -8.55 & 4.20 & 0.1238 & 0.03328 & HIRES \\
2458828.13122 & 12.10 & 4.06 & 0.1250 & 0.03326 & HIRES \\
2458833.12337 & -9.99 & 4.52 & 0.1246 & 0.03333 & HIRES \\
2458833.12694 & -4.73 & 4.81 & 0.1229 & 0.03317 & HIRES \\
2458833.13057 & -3.89 & 4.53 & 0.1250 & 0.03330 & HIRES \\
2458834.06001 & -4.62 & 4.23 & 0.1254 & 0.03302 & HIRES \\
2458834.06378 & -11.58 & 4.15 & 0.1245 & 0.03303 & HIRES \\
2458834.06782 & -9.07 & 3.94 & 0.1255 & 0.03295 & HIRES \\
2458834.15223 & -13.96 & 3.92 & 0.1236 & 0.03314 & HIRES \\
2458834.15545 & -17.50 & 4.11 & 0.1242 & 0.03304 & HIRES \\
2458834.15878 & -21.43 & 4.21 & 0.1235 & 0.03304 & HIRES \\
2458878.94352 & -3.54 & 4.98 & 0.1250 & 0.03284 & HIRES \\
2458879.94452 & -4.14 & 4.50 & 0.1242 & 0.03284 & HIRES \\
2458881.04737 & -8.74 & 4.47 & 0.1237 & 0.03324 & HIRES \\
2458881.05080 & 8.23 & 4.21 & 0.1242 & 0.03307 & HIRES \\
2458881.05422 & 6.24 & 4.50 & 0.1239 & 0.03313 & HIRES \\
2457759.80567 & 26.08 & 6.79 & 0.1515 & -- & PFS \\
2457761.81934 & -2.24 & 7.28 & 0.1556 & -- & PFS \\
2457763.85691 & -3.61 & 6.75 & 0.1537 & -- & PFS \\
2457765.86413 & -9.28 & 7.23 & 0.1512 & -- & PFS \\
2457767.85472 & -11.72 & 6.54 & 0.1515 & -- & PFS \\
2457769.84375 & 1.40 & 8.21 & 0.2536 & -- & PFS \\
2458207.69293 & 8.41 & 6.81 & 0.1614 & -- & PFS \\
2458265.54169 & 8.55 & 5.58 & 0.1696 & -- & PFS \\
2458265.60975 & 0.00 & 5.03 & 0.1669 & -- & PFS \\
2458266.54613 & 0.21 & 5.40 & 0.1691 & -- & PFS \\
2458266.63963 & 6.08 & 5.00 & 0.1803 & -- & PFS \\
2458270.49736 & 14.74 & 4.90 & 0.1594 & -- & PFS \\
2458270.65094 & 8.27 & 4.74 & 0.1737 & -- & PFS \\
2458271.53219 & 4.71 & 4.53 & 0.1563 & -- & PFS \\
2458271.62931 & 1.16 & 4.64 & 0.1614 & -- & PFS \\
2458272.49145 & 6.07 & 4.65 & 0.1687 & -- & PFS \\
2458272.50365 & -15.94 & 5.76 & 0.1867 & -- & PFS \\
2458272.60086 & -3.55 & 4.61 & 0.1573 & -- & PFS \\
2458292.49251 & -3.61 & 4.53 & 0.1599 & -- & PFS \\
2458292.56916 & -7.46 & 5.09 & 0.1550 & -- & PFS \\
2458294.48588 & -2.15 & 4.84 & 0.1627 & -- & PFS \\
2458294.56182 & -17.19 & 5.39 & 0.1649 & -- & PFS \\
2458296.51344 & -12.01 & 4.90 & 0.1542 & -- & PFS \\
2458299.51510 & -6.25 & 4.93 & 0.1607 & -- & PFS \\
2457781.06111 & -5.04 & 13.21 & 0.1291 & -- & APF \\
2457809.02734 & -30.19 & 22.39 & 0.1200 & -- & APF \\
2457809.05593 & -6.34 & 17.44 & 0.1431 & -- & APF \\
2457815.06470 & 14.43 & 12.31 & 0.1279 & -- & APF \\
2457865.81600 & -2.68 & 11.12 & 0.1335 & -- & APF \\
2457809.04177 & -38.47 & 18.68 & 0.1239 & -- & APF \\
2457896.72618 & -1.23 & 14.79 & 0.1283 & -- & APF \\
2457873.86098 & 0.96 & 11.55 & 0.1295 & -- & APF \\
2457815.05044 & 30.82 & 11.67 & 0.1247 & -- & APF \\
2457901.75498 & -3.58 & 11.65 & 0.1335 & -- & APF \\
2457882.81860 & -44.66 & 13.42 & 0.1296 & -- & APF \\
2457796.97958 & -5.63 & 17.80 & 0.1282 & -- & APF \\
2457780.91058 & 48.59 & 18.83 & 0.1266 & -- & APF \\
2457821.89386 & 141.13 & 18.37 & 0.1297 & -- & APF \\
2457822.00237 & -5.56 & 16.55 & 0.1204 & -- & APF \\
2457897.73030 & -16.58 & 12.26 & 0.1311 & -- & APF \\
2457752.06664 & 11.36 & 10.45 & 0.1392 & -- & APF \\
2457848.82814 & -47.54 & 24.27 & 0.1367 & -- & APF \\
2457814.83360 & -8.39 & 11.33 & 0.1328 & -- & APF \\
2457896.74040 & -18.95 & 12.14 & 0.1271 & -- & APF \\
2457816.04258 & 29.95 & 11.33 & 0.1302 & -- & APF \\
2457780.96595 & -5.23 & 12.58 & 0.1261 & -- & APF \\
2457873.88982 & -20.07 & 13.26 & 0.1350 & -- & APF \\
2457816.05647 & -3.25 & 12.08 & 0.1615 & -- & APF \\
2457781.03551 & 29.69 & 15.55 & 0.1342 & -- & APF \\
2457893.73144 & -12.69 & 11.80 & 0.1297 & -- & APF \\
2457893.78716 & -21.05 & 12.27 & 0.1311 & -- & APF \\
2457796.91448 & 5.30 & 21.96 & 0.1281 & -- & APF \\
2457873.75443 & 19.37 & 12.28 & 0.1293 & -- & APF \\
2457781.10700 & -42.78 & 16.88 & 0.1438 & -- & APF \\
2457882.75310 & 17.77 & 12.22 & 0.1531 & -- & APF \\
2457783.84848 & -24.83 & 15.64 & 0.1158 & -- & APF \\
2457893.80928 & -158.14 & 42.60 & -- & -- & APF \\
2457750.09575 & -3.35 & 11.74 & 0.1271 & -- & APF \\
2457845.84002 & -37.50 & 12.19 & 0.1267 & -- & APF \\
2457901.78421 & 17.07 & 11.74 & 0.1292 & -- & APF \\
2457780.91969 & 12.76 & 17.97 & 0.1264 & -- & APF \\
2457848.79785 & -69.33 & 24.00 & 0.2515 & -- & APF \\
2457783.87896 & -14.96 & 14.39 & 0.1374 & -- & APF \\
2457814.93242 & 2.54 & 10.07 & 0.1320 & -- & APF \\
2457823.76616 & 87.35 & 14.36 & 0.1375 & -- & APF \\
2457815.03593 & 12.47 & 11.61 & 0.1320 & -- & APF \\
2457847.93237 & -61.36 & 11.48 & 0.1262 & -- & APF \\
2457784.00396 & -1.34 & 11.82 & 0.1244 & -- & APF \\
2457847.94652 & -16.56 & 13.23 & 0.1318 & -- & APF \\
2457901.76933 & -16.30 & 11.76 & 0.1263 & -- & APF \\
2457783.98895 & 15.76 & 12.71 & 0.1350 & -- & APF \\
2457752.03791 & -2.96 & 10.43 & 0.1350 & -- & APF \\
2457845.98544 & -28.47 & 13.65 & 0.1396 & -- & APF \\
2457893.79993 & -8.99 & 12.85 & 0.1385 & -- & APF \\
2457780.90075 & 0.82 & 20.01 & 0.1329 & -- & APF \\
2457784.10953 & -81.15 & 15.50 & 0.1450 & -- & APF \\
2457750.06719 & 52.66 & 12.77 & 0.1296 & -- & APF \\
2457781.04973 & -3.73 & 14.98 & 0.1348 & -- & APF \\
2457865.80733 & -19.26 & 22.07 & 0.1329 & -- & APF \\
2457847.91817 & -20.83 & 11.73 & 0.1332 & -- & APF \\
2457873.74045 & -2.22 & 11.53 & 0.1345 & -- & APF \\
2457752.05226 & 25.85 & 9.94 & 0.1328 & -- & APF \\
2457815.84613 & 7.36 & 11.11 & 0.1281 & -- & APF \\
2457823.84886 & 81.62 & 14.50 & 0.1267 & -- & APF \\
2457882.74019 & -41.25 & 11.84 & 0.1340 & -- & APF \\
2457814.81938 & -6.62 & 11.75 & 0.1295 & -- & APF \\
2457877.83994 & -26.05 & 12.92 & 0.1327 & -- & APF \\
2457894.74722 & 2.66 & 11.74 & 0.1363 & -- & APF \\
2457882.84429 & -22.37 & 13.67 & 0.1335 & -- & APF \\
2457919.74774 & 9.62 & 12.86 & 0.1360 & -- & APF \\
2457873.76897 & -18.00 & 13.05 & 0.1306 & -- & APF \\
2457783.86216 & 0.61 & 14.70 & 0.1292 & -- & APF \\
2457893.74604 & -5.52 & 11.15 & 0.1250 & -- & APF \\
2457897.71559 & -10.66 & 11.75 & 0.1322 & -- & APF \\
2457823.75089 & 63.51 & 15.59 & 0.1228 & -- & APF \\
2457865.83042 & 1.38 & 12.02 & 0.1344 & -- & APF \\
2457822.01596 & 72.13 & 18.72 & 0.1306 & -- & APF \\
2457796.92414 & -77.42 & 21.74 & 0.1135 & -- & APF \\
2457780.95155 & 1.04 & 12.42 & 0.1285 & -- & APF \\
2457877.78376 & 5.38 & 11.67 & 0.1306 & -- & APF \\
2457894.71857 & -5.35 & 11.55 & 0.1319 & -- & APF \\
2457815.91674 & 2.57 & 11.42 & 0.1319 & -- & APF \\
2457814.84816 & -15.17 & 11.87 & 0.1258 & -- & APF \\
2457877.75768 & -52.78 & 11.60 & 0.1270 & -- & APF \\
2457815.81820 & -2.70 & 11.36 & 0.1320 & -- & APF \\
2457784.01797 & 3.61 & 13.70 & 0.1309 & -- & APF \\
2457873.87517 & -17.82 & 13.27 & 0.1263 & -- & APF \\
2457845.85385 & -17.22 & 13.75 & 0.1340 & -- & APF \\
2457822.03064 & 72.96 & 18.27 & 0.1234 & -- & APF \\
2457823.73526 & 158.58 & 20.04 & 0.1148 & -- & APF \\
2457887.79156 & 10.47 & 21.52 & 0.1353 & -- & APF \\
2457823.83544 & 122.21 & 14.40 & 0.1289 & -- & APF \\
2457750.08117 & 8.69 & 12.42 & 0.1326 & -- & APF \\
2457815.93185 & -39.58 & 11.79 & 0.1243 & -- & APF \\
2457919.77509 & 15.30 & 13.33 & 0.1288 & -- & APF \\
2457877.85293 & 25.49 & 12.10 & 0.1319 & -- & APF \\
2457877.77163 & -12.37 & 11.11 & 0.1318 & -- & APF \\
2457815.83197 & 21.74 & 11.47 & 0.1270 & -- & APF \\
2457896.71402 & -14.85 & 12.71 & 0.1296 & -- & APF \\
2457893.71726 & 3.00 & 10.32 & 0.1307 & -- & APF \\
2457823.82109 & 50.03 & 13.31 & 0.1279 & -- & APF \\
2457796.90531 & -24.95 & 21.01 & 0.1225 & -- & APF \\
2457887.81092 & -7.28 & 15.47 & 0.1321 & -- & APF \\
2457816.07010 & 24.14 & 13.86 & 0.1306 & -- & APF \\
2457796.98993 & -36.61 & 17.39 & 0.1185 & -- & APF \\
2457829.94568 & -14.07 & 10.74 & 0.1330 & -- & APF \\
2457784.09563 & -4.87 & 12.95 & 0.1541 & -- & APF \\
2457796.96897 & 30.46 & 18.35 & 0.1279 & -- & APF \\
2457845.86927 & -31.55 & 13.66 & 0.1339 & -- & APF \\
2457882.76205 & 9.55 & 16.04 & -- & -- & APF \\
2457821.87919 & 50.01 & 16.52 & 0.1379 & -- & APF \\
2457814.91859 & -13.03 & 9.80 & 0.1325 & -- & APF \\
2457894.80200 & 16.97 & 11.99 & 0.1304 & -- & APF \\
2457894.81491 & 15.35 & 12.04 & 0.1308 & -- & APF \\
2457894.73280 & -12.60 & 10.89 & 0.1350 & -- & APF \\
2457829.96052 & -11.24 & 11.38 & 0.1287 & -- & APF \\
2457829.97503 & -2.62 & 11.61 & 0.1351 & -- & APF \\
2457897.74280 & 6.85 & 12.72 & 0.1250 & -- & APF \\
2457780.98023 & 13.16 & 12.70 & 0.1332 & -- & APF \\
2457919.76193 & 3.86 & 12.00 & 0.1334 & -- & APF \\
2457815.90276 & 15.11 & 10.73 & 0.1280 & -- & APF \\
2457887.80079 & -10.67 & 20.01 & 0.1378 & -- & APF \\
2457845.95683 & -27.37 & 12.47 & 0.1294 & -- & APF \\
2457882.83131 & -38.41 & 14.90 & 0.1309 & -- & APF \\
2457821.86260 & 44.47 & 15.33 & 0.1332 & -- & APF \\
2457848.81036 & -130.55 & 25.27 & 0.1204 & -- & APF \\
2457814.94576 & 1.03 & 9.37 & 0.1291 & -- & APF \\
2457845.97122 & -38.39 & 12.72 & 0.1342 & -- & APF \\
2457894.78750 & 8.53 & 10.56 & 0.1274 & -- & APF
\enddata
\tablenotetext{}{S$_{\rm HK}$ values have an uncertainty of 0.001 for HIRES data, 0.002 for APF data, and no calculated uncertainties for PFS data.}
\end{deluxetable*}

\startlongtable
\begin{deluxetable*}{cc}
\tablecaption{\gj\ Photometry \label{tab:gjphot}}
\tablehead{
 \colhead{Time (HJD)} & \colhead{Differential Magnitude (Cousins R)}
}
\startdata
2458384.8398 & -3.29860 \\
2458387.7886 & -3.29043 \\
2458388.7836 & -3.29695 \\
2458389.7598 & -3.29301 \\
2458390.7934 & -3.29492 \\
2458395.8109 & -3.29757 \\
2458396.7518 & -3.29810 \\
2458397.7672 & -3.29389 \\
2458401.7730 & -3.29437 \\
2458409.7743 & -3.29824 \\
2458411.7562 & -3.29593 \\
2458416.7611 & -3.29737 \\
2458417.7244 & -3.29568 \\
2458418.7359 & -3.29437 \\
2458419.7340 & -3.29448 \\
2458420.7310 & -3.29879 \\
2458424.7217 & -3.30008 \\
2458425.7338 & -3.28865 \\
2458426.7161 & -3.29505 \\
2458428.7217 & -3.29477 \\
2458429.7013 & -3.29479 \\
2458430.7246 & -3.29014 \\
2458431.7200 & -3.29774 \\
2458432.7074 & -3.29444 \\
2458433.6980 & -3.29743 \\
2458434.6915 & -3.29221 \\
2458435.7153 & -3.29915 \\
2458438.6970 & -3.30280 \\
2458439.7007 & -3.29541 \\
2458442.7045 & -3.30265 \\
2458443.7110 & -3.29193 \\
2458447.7030 & -3.28811 \\
2458448.6802 & -3.29624 \\
2458456.6973 & -3.29770 \\
2458462.6936 & -3.29643 \\
2458465.6939 & -3.29154 \\
2458466.6406 & -3.29506 \\
2458472.6551 & -3.29889 \\
2458473.6821 & -3.29741 \\
2458477.6708 & -3.29311 \\
2458487.6505 & -3.30077 \\
2458488.6392 & -3.29675 \\
2458510.5919 & -3.29905 \\
2458630.9510 & -3.29773 \\
2458635.9774 & -3.29121 \\
2458639.9574 & -3.28977 \\
2458640.9686 & -3.29189 \\
2458641.9467 & -3.28610 \\
2458654.9693 & -3.30466 \\
2458655.9418 & -3.29611 \\
2458657.9366 & -3.29868 \\
2458757.7474 & -3.29591 \\
2458762.7384 & -3.29828 \\
2458763.8894 & -3.29123 \\
2458765.7077 & -3.30108 \\
2458766.7137 & -3.29908 \\
2458770.7195 & -3.29695 \\
2458771.6987 & -3.29525 \\
2458775.7327 & -3.29347 \\
2458777.7144 & -3.29992 \\
2458778.6961 & -3.29596 \\
2458780.6850 & -3.29421 \\
2458781.7094 & -3.28570 \\
2458784.6903 & -3.29402 \\
2458787.6802 & -3.29677 \\
2458788.7024 & -3.29800 \\
2458800.7053 & -3.29736 \\
2458801.6652 & -3.30072 \\
2458802.6866 & -3.29158 \\
2458831.7012 & -3.30000 \\
2458833.6352 & -3.29162 \\
2458860.6088 & -3.29597 \\
2458867.5926 & -3.29273 \\
2458876.5765 & -3.29225 
\enddata
\end{deluxetable*}

\startlongtable
\begin{deluxetable*}{cc}
\tablecaption{\hd\ Photometry \label{tab:hdphot}}
\tablehead{
 \colhead{Time (HJD)} & \colhead{Differential Magnitude ((b+y)/2)}
}
\startdata
2458159.9462 & 1.58370 \\
2458161.9455 & 1.58287 \\
2458162.8071 & 1.58217 \\
2458172.7803 & 1.58193 \\
2458174.7746 & 1.58083 \\
2458174.9086 & 1.58407 \\
2458176.8481 & 1.58453 \\
2458183.8548 & 1.58830 \\
2458184.7545 & 1.58647 \\
2458184.8741 & 1.58447 \\
2458189.7523 & 1.58600 \\
2458195.7926 & 1.58200 \\
2458195.8678 & 1.58800 \\
2458197.7836 & 1.58283 \\
2458197.8585 & 1.58413 \\
2458204.7111 & 1.58447 \\
2458204.8259 & 1.58257 \\
2458205.7376 & 1.58827 \\
2458205.7839 & 1.58210 \\
2458210.7527 & 1.58693 \\
2458210.8097 & 1.58093 \\
2458211.7425 & 1.58367 \\
2458212.7486 & 1.58580 \\
2458212.8103 & 1.58507 \\
2458213.7943 & 1.58350 \\
2458216.7387 & 1.58667 \\
2458218.7464 & 1.58583 \\
2458220.8272 & 1.59000 \\
2458223.7525 & 1.58690 \\
2458228.7301 & 1.58970 \\
2458229.7351 & 1.58503 \\
2458231.6962 & 1.58663 \\
2458241.7346 & 1.58417 \\
2458242.7333 & 1.58747 \\
2458250.7291 & 1.58127 \\
2458256.7137 & 1.58497 \\
2458257.7085 & 1.58543 \\
2458258.6988 & 1.58087 \\
2458266.6928 & 1.59070 \\
2458267.7002 & 1.58433 \\
2458272.6969 & 1.58287 \\
2458273.6955 & 1.58683 \\
2458277.6920 & 1.58887 \\
\enddata
\end{deluxetable*}

\acknowledgments
We thank the anonymous reviewer for their time and helpful comments.

M.R.K is supported by the NSF Graduate Research Fellowship, grant No. DGE 1339067.

C.P. is supported by the Technologies for Exo-Planetary Science (TEPS) CREATE program and further acknowledges financial support by the Fonds de Recherche Qu\'{e}b\'{e}cois—Nature et Technologie (FRQNT; Qu\'{e}bec).

G.W.H. acknowledges long-term support from NASA, NSF, Tennessee State University, and the State of Tennessee through its Centers of Excellence program.

L.M.W. is supported by the Beatrice Watson Parrent Fellowship and NASA ADAP Grant 80NSSC19K0597. 

P.D. acknowledges support from a National Science Foundation Astronomy and Astrophysics Postdoctoral Fellowship under award AST-1903811.

JMAM gratefully acknowledges support from the National Science Foundation Graduate Research Fellowship Program under Grant No. DGE-1842400. JMAM also thanks the LSSTC Data Science Fellowship Program, which is funded by LSSTC, NSF Cybertraining Grant No. 1829740, the Brinson Foundation, and the Moore Foundation; his participation in the program has benefited this work.

The authors wish to recognize and acknowledge the very significant cultural role and reverence that the summit of Maunakea has always had within the indigenous Hawaiian community. We are most fortunate to have the opportunity to conduct observations from this mountain.

Some of the observations in the paper made use of the High-Resolution Imaging instrument Zorro. Zorro was funded by the NASA Exoplanet Exploration Program and built at the NASA Ames Research Center by Steve B. Howell, Nic Scott, Elliott P. Horch, and Emmett Quigley. Zorro is mounted on the Gemini South telescope of the international Gemini Observatory, a program of NSF’s OIR Lab, which is managed by the Association of Universities for Research in Astronomy (AURA) under a cooperative agreement with the National Science Foundation. 

This work is based in part on observations made with the {\it Spitzer} Space Telescope, which was operated by the Jet Propulsion Laboratory, California Institute of Technology under a contract with NASA. Support for this work was provided by NASA through an award issued by JPL/Caltech.

This research has made use of the Exoplanet Follow-up Observing Program (ExoFOP), which is operated by the California Institute of Technology, under contract with the National Aeronautics and Space Administration.

Part of the research was carried out at the Jet Propulsion Laboratory, California Institute of Technology, under a contract with the National Aeronautics and Space Administration (80NM0018D0004).

This work has made use of data from the European Space Agency (ESA) mission
{\it Gaia} (\url{https://www.cosmos.esa.int/gaia}), processed by the {\it Gaia}
Data Processing and Analysis Consortium (DPAC,
\url{https://www.cosmos.esa.int/web/gaia/dpac/consortium}). Funding for the DPAC
has been provided by national institutions, in particular the institutions
participating in the {\it Gaia} Multilateral Agreement.

\vspace{5mm}
\facilities{Keck:I(HIRES), Magellan:Clay(PFS), Spitzer, APF, TSU:AIT, Gemini:South(Zorro)}

\software{\texttt{radvel} \citep{Fulton2018}, \texttt{batman} \citep{Kreidberg2015}, \texttt{SpecMatch-Emp} \citep{Yee2017}, \texttt{isoclassify} \citep{Huber2017}, \texttt{spock} \citep{Tamayo2020}, \texttt{rebound} \citep{Rein2011}, \texttt{numpy} \citep{vanderwalt2011}, \texttt{astropy} \citep{Astropy2013}, \texttt{emcee} \citep{ForemanMackey2013}}

\bibliography{main}{}
\bibliographystyle{aasjournal}



\end{document}